\documentclass{article}

\usepackage{graphicx}  
\usepackage{amsmath}   
\usepackage[compress]{cite}
\usepackage{amssymb}   

\usepackage[active]{srcltx}

\usepackage{color}

\def\eq#1{{Eq.~(\ref{#1})}}

\def\cc{{cosmological\ constant}}

\newcommand{\LL}{Lanczos-Lovelock }

\def\md{microscopic degrees of freedom of the spacetime}

\title{The Atoms Of Space, Gravity and the Cosmological Constant}
 
\author{T. Padmanabhan\\
IUCAA, Pune University Campus, \\
Ganeshkhind, Pune 411007, India.\\
email: paddy@iucaa.in
}

\date{ }

\begin{document}

\maketitle

\begin{abstract}
I describe an approach which connects  classical gravity with the quantum microstructure of spacetime. The field equations arise from maximizing the density of states of  matter plus geometry. The former is identified using the thermodynamics of null surfaces while the latter arises due to the existence of a zero-point length in the spacetime. The resulting field equations remain invariant when a constant is added to the matter Lagrangian, which is a symmetry of the matter sector. Therefore, the \cc\ arises as an integration constant. A non-zero value $(\Lambda)$ of the \cc\ renders the amount of cosmic information $(I_c)$ accessible to an eternal observer finite and hence is directly related to it. This relation allows us to determine the numerical value of $(\Lambda)$ from the quantum structure of spacetime.
\end{abstract}



\tableofcontents

\section{Gravity from the atoms of space: Summary}
\label{sec:grfratoms}

I describe  an approach which obtains the gravitational field equations from a thermodynamic variational principle and, as a bonus, allows you to determine the numerical value of the \cc. This variational principle can be interpreted in terms of the number density of \md\ which I will call, figuratively, as the `atoms of space'. It turns out that one can make a significant amount of progress in motivating and understanding the nature of these atoms of space in a ``top-down'' approach, starting from known properties of classical gravity and the thermodynamics of null surfaces. 
This works out, essentially, because the horizons act as magnifying glasses\cite{tplens} for Planck scale physics through the large redshifts they introduce.

The ``top-down'' route --- from classical gravity to quantum structure of spacetime --- is what I will follow from the next section onwards. But it is useful to summarize the ``bottom-up'' picture which emerges from this description so that the broader perspective  remains in focus. The purpose of this introductory section is to do this, postponing the details of arguments and mathematical proofs to later sections. 

A fundamental quantity in the description of, say,  a fluid is the distribution function $f(x^i, p_j)$ which \textit{counts}  the number of  atoms $dN = f(x^i, p_j)d^3xd^3p$   per unit phase space volume $d^3xd^3p$. (The norm of the four-momentum $p^2 $ is fixed by the mass of the particles, making the invariant phase space volume six dimensional.)\footnote{I use the signature $(-,+,+,+)$ and natural units with $c=1, \hbar =1$ and set $\kappa = 8\pi G =8\pi L_P^2$ where $L_P$ is the Planck length $(G\hbar/c^3)^{1/2}$ in natural units.  Latin letters $i, j$ etc. range over spacetime indices and the Greek letters $\alpha, \beta$ etc. range over the spatial indices.}
 The description in terms of a distribution function is remarkable because it allows us to use the continuum language and --- at the same time --- recognize the discrete nature of the fluid.\footnote{One could equivalently think of $f$ as the number of \textit{degrees of freedom} per unit phase space volume or even as the \textit{probability} of occupation of a given phase space volume. It will be conceptually convenient to switch between these equivalent descriptions whenever appropriate.}

In a similar manner, I want to describe the spacetime as a fluid containing the atoms of space described by a number density of \md, denoted by
$\rho (x,\phi_A)$. Here, $\phi_A$ (with $A=1,2,3,...$) denotes possible internal degrees of freedom (analogous to the momentum $p_i$ for the distribution function for the molecules of a fluid).  The dependence on $x^i$ arises only indirectly through  (i) the geometrical variables like the metric tensor, curvature tensor etc., (which I will collectively denote as $\mathcal{G}_N(x)$ with $N=1,2,3,...$) and (ii) the matter sector described by a symmetric divergence-free energy momentum tensor $T_{ab}(x)$; so $\rho (x,\phi_A)=\rho [\mathcal{G}_N (x),\phi_A, T_{ab}(x)]$. The total number of degrees of freedom for a given spacetime configuration is given by the product over all phase space: 
\begin{equation}
 \Omega_{\rm tot} =\prod_{\phi_A}\ \prod_{x}\, \rho[\mathcal{G}_N (x),\phi_A, T_{ab}(x)] \equiv \exp S_{\rm tot}
\end{equation} 
My aim is to obtain the gravitational field equations by maximizing  the expression for $\Omega_{\rm tot}$ or, equivalently, the configurational entropy $S_{\rm tot}$. 

It will turn out that $\rho$ can  be expressed as a product $\rho=\rho_g (\mathcal{G}_N(x), \phi_A)\, \rho_m (T_{ab}(x), \phi_A)$ where $\rho_g (\mathcal{G}_N(x), \phi_A))$
represents the spacetime contribution and $\rho_m (T_{ab}(x), \phi_A)$ describes the effect of matter. (What is relevant, of course, is the product,  $\rho_g\rho_m$; but it is conceptually useful to separate them as the product of two factors.)
The $\Omega_{\rm tot}$ becomes:
\begin{equation}
 \Omega_{\rm tot} =\prod_{\phi_A}\ \prod_{x}\, \rho_g (\mathcal{G}_N, \phi_A)\, \rho_m (T_{ab}, \phi_A) \equiv \exp [S_{\rm grav}+S_{\rm m}]
\end{equation} 
To obtain the classical limit of gravity, it is convenient to leave the product over the internal variable $\phi_A$ as it is and exponentiate the product over $x$, thereby obtaining 
\begin{equation}
 \Omega_{\rm tot} =\prod_{\phi_A} \exp \sum_x \left( \ln \rho_g + \ln \rho_m\right)
 \label{rt2}
\end{equation}
It turns out that the saddle point maximum of the expression within the brackets in \eq{rt2} with respect to $\phi_A$ reproduces the Einstein's equations for gravity.

Obviously, this  result depends on the  expressions for $\rho_g, \rho_m$ and the internal variable $\phi_A$. I will show (see Sec. \ref{subsec:matter} and Sec. \ref{subsec:geometry}) that the internal variable $\phi_A$ can be mapped to a four-vector $n^a$ of constant norm. (Its norm is unity in the Euclidean sector and it will map to a zero norm null vector in the Lorentzian sector.) So, 
\begin{equation}
 \rho_g(x,\phi_A) = \rho_g (x^i,n_j) = \rho_g (t, \mathbf{x}; \mathbf{n})
\end{equation} 
is completely analogous to the distribution function for zero mass particles (i.e., a null fluid) in the spacetime. In terms of this vector field $n^a$, the $\rho_m$ and $\rho_g$ are given, at the leading order, by the expressions
\begin{equation}
\ln \rho_m \equiv L_P^4\mathcal{H}_m = L_P^4 T_{ab} n^a n^b; \qquad \ln \rho_g \approx - \frac{L_P^2}{8\pi} R_{ab} n^an^b
\label{rt4}
\end{equation}
I will derive these expressions in Sec. \ref{subsec:matter} and Sec. \ref{sec:zpl} respectively. Using these expressions in \eq{rt2}, we see that the expression in the square bracket reduces to one proportional to 
\begin{equation}
E^a_b n_an^b\equiv \left(T^a_b (x) - \frac{1}{\kappa} R^a_b (x)\right) n_an^b 
\label{tponen}
\end{equation} 
The extremum condition for this expression, with respect   to $n^a \to n^a + \delta n^a$, 
subject to the constraint $n^2=$ constant, leads to Einstein's equations, with a cosmological constant arising as an integration constant. (This should be obvious. I will describe a somewhat more general result in Sec. \ref{sec:vpg})
It will turn out that, in the classical limit, $E^a_b n_a n^b$ can indeed be interpreted as the rate of heating per unit area of a null surface, thereby making this a thermodynamical variational principle, for the fluid we call spacetime.

This alternate perspective resonates well with several peculiar features of gravity, especially the connection with horizon thermodynamics. I will also show that this approach offers fresh insights into the \cc\ problem and, in fact, allows us to determine its numerical value quite accurately.  It is therefore difficult to ignore the alternative insights provided by this approach.

\section{Three avatars of gravitational field equations}
\label{sec:classgrav}

I shall now describe the details of this formalism starting from classical gravity and working towards deeper layers.

The field equation in Einstein's gravity is usually expressed in terms of $G_{ab} \equiv R_{ab} - (1/2)g_{ab}R$ in the form
\begin{equation}
 G_{ab} = \kappa T_{ab} 
\label{ee1}
\end{equation} 
This is what you learn in standard textbooks. But there are  two other ---  and as I will argue,  nicer --- ways of writing the gravitational field equation.

The first alternative is to introduce a timelike, normalized vector field $u^i$ (which could be thought of as the four-velocity of a fiducial observer) and demand that the equation 
\begin{equation}
 G_{ab}u^au^b = \kappa T_{ab}u^au^b 
\label{ee2}
\end{equation} 
holds for \textit{all} observers. This demand, of course, can be met only if \eq{ee1} holds and we recover the standard result. The second alternative is to introduce a \textit{null} vector field $\ell^a$ (which could be thought of as a normal to  a null surface in the spacetime) and demand that the equation 
\begin{equation}
 G_{ab}\ell^a \ell^b = R_{ab}\ell^a \ell^b = \kappa T_{ab}\ell^a \ell^b
\label{ee3}
\end{equation}
holds for all null vectors $\ell^a$.  This leads to the result
\begin{equation}
 G_{ab} = \kappa T_{ab}  + \Lambda g_{ab}
\label{ee4}
\end{equation}
where $\Lambda$ is a constant.\footnote{Equation~(\ref{ee3}) implies $R^{a}_{b} - \kappa T^a_{b} = f(x) \delta^a_{b}$. Taking the divergence and using the facts that $\nabla_a T^a_b =0$ and 
$\nabla_a R^a_b = (1/2)\partial_b R$ tells you that $f(x) = (1/2)R +$ a constant, leading to \eq{ee4}. It is sometimes claimed that the Bianchi identity $\nabla_aG^a_b =0$ \textit{implies} $\nabla_aT^a_b =0$. But  $T^a_b $ can be defined through the variation of the matter Lagrangian with respect to arbitrary coordinate transformations $x^a\to x^a+\xi^a(x)$. Its conservation, $\partial_a T^a_b =0$, in Cartesian coordinates in local inertial frames, becomes $\nabla_aT^a_b =0$ in curvilinear coordinates. The principle of equivalence now demands the validity of this condition in an arbitrary curved spacetime. 
That is, you can \textit{derive}  $\nabla_a T^a_b =0$ without using the Bianchi identity or the field equations.
It is, therefore, more appropriate to think of the Bianchi identity as \textit{being consistent} with $\nabla_aT^a_b =0$ rather than \textit{implying} it.} 
So, \eq{ee3} also leads to  \eq{ee1} but with a crucial difference: It allows for a cosmological constant $\Lambda$ to arise as an integration constant to the field equation.

While all the three formulations are algebraically equivalent, they are conceptually very different.  Two such differences  between \eq{ee1} and either of \eq{ee2} or \eq{ee3} are immediately noticeable. First, \eq{ee2} and \eq{ee3} involve additional vector fields but are scalar equations. They contain the same information content as the ten tensor components of \eq{ee1} because we demand them to hold for \textit{all} $u^i$ or all $\ell^i$. If you think of $u^i$ as a four velocity of an observer, then \eq{ee2} is a statement about the equality of two quantities which this observer  measures in the matter sector and the geometrical sector. Such a statement, invoking a class of observers, is similar in spirit to the way we obtain the kinematics of gravity (``how gravity makes matter move'') by introducing special relativity in the coordinate frames adapted to the freely falling observers.

Second, nobody has come up with a physical meaning for the text book field equation, expressed in the form \eq{ee1}. The right hand side, of course,  has the physical meaning as the energy momentum tensor but not the left hand side. 
In the conventional approach,  we actually do not have a \textit{mechanism} which tells us how $T^a_b$ ends up curving the spacetime. The relation  $G^a_b = \kappa T^a_b$  equates apples and oranges; the left hand side is purely geometrical while the right hand side is made of  matter which we know has a large number of discrete (quantum) degrees of freedom.\footnote{The usual approach is to use the quantum expectation value $\langle T^a_b\rangle$ in the right hand side but that is hardly appropriate as a fundamental description and provides us with no useful insights.} An equation of the kind, $G^a_bn_a n^b = \kappa  T^a_bn_a n^b$ (where $n_a = u_a$ or $\ell_a$), on the other hand, is conceptually better in this regard.  We can hope to interpret both sides independently  and think of this equation as a balancing act performed by spacetime. As we shall see, a description is reinforced by the extremum principle in which both $R^a_b n_a n^b$ and $ T^a_bn_a n^b$ can be thought of as distorting the value of $\rho_g\rho_m$ from unity, with the gravitational field equations restoring the value $\rho_g\rho_m=1$ on-shell. 
Much of the later sections of this article will be devoted to  providing the physical meaning for the left hand side of \eq{ee3} which will turn out to be  thermodynamic in nature. In short, \eq{ee2} and \eq{ee3} possess nicer physical interpretations than \eq{ee1}.

\subsection{What Einstein could have done!}
\label{subsec:ae}

Before proceeding further, I will provide a straightforward derivation and interpretation of \eq{ee2} showing how Einstein could have obtained this equation instead of \eq{ee1} (thereby saving us a lot of trouble!).

What Einstein was looking for was a generalization of the field equation for gravity in Newtonian theory.
The dynamics of Newtonian gravity is governed by the Poisson equation $\nabla^2 (2\phi) =  \kappa \rho$ which relates the gravitational potential $\phi$ to the mass density $\rho$. When we move on to general relativity, the principle of equivalence identifies the gravitational potential with a component of the metric tensor through $g_{00} = - (1+2\phi)$ so that the Poisson equation can be formally written as  $-\nabla^2 g_{00} = \kappa T_{00}$ where $\rho$ is identified with the time-time component $T_{00}$ of the divergence-free, second rank symmetric energy momentum tensor $T_{ab}$. 
Since $\nabla^2$ is not Lorentz invariant, one might think (alas, wrongly!) that it is preferable to  ``generalize'' the $\nabla^2$ to $\square^2$ so that the left hand side has second derivatives in \textit{both space and time}.  
The second derivatives of the metric tensor can be expressed covariantly  in terms of the curvature tensor, which led Einstein to look for a   divergence-free, second rank symmetric tensor to replace $\nabla^2 g_{00}$ in the left hand side. After several false starts, he came up with \eq{ee1}  and postulated it to be the field equation.

But Einstein could have taken a different, and better, route! 
One can come up with a relativistic generalization of Newton's law of gravity $\nabla^2\phi\propto\rho$, \textit{retaining the right hand side as it is and without introducing second time derivatives in the left hand side}. 

To do this, notice that: 
(i) The energy density  $\rho=T_{ab}u^au^b$, which appears in the right hand side, is foliation/observer dependent where $u^i$ is the four velocity of an observer. There is no way you can keep $u^i$ out of it and you should accept it as a fact of life. 
(ii) Since $g_{ab}$ plays the role of $\phi/c^2$, a covariant scalar which generalizes  the left hand side, $\nabla^2\phi$, could indeed come from the curvature tensor --- which contains the second derivatives of the metric. 
But, you need to find a generalization  \textit{which depends on the four-velocity $u^i$ of the observer} because the right hand side does. A purely geometrical object (like e.g. $R$), won't do.
(iii) It is, of course, perfectly acceptable for the left hand side \textit{not} to have second \textit{time} derivatives of the metric, in the rest frame of the observer, since they do not occur in $\nabla^2\phi$. 

To obtain a scalar with  \textit{spatial} second derivatives which depends on $u^i$ (to replace $\nabla^2\phi$),  we first project the indices of $R_{abcd}$ to the space orthogonal to $u^i$, using the projection tensor $P^i_j=\delta^i_j+u^iu_j$, thereby obtaining the tensor
$\mathcal{R}_{ijkl}\equiv P^a_iP^b_jP^c_kP^d_l R_{abcd}$. The only scalar we can construct from $\mathcal{R}_{ijkl}$ is $\mathcal{R}^{-2}\equiv\mathcal{R}_{ij}^{ij}$ where $\mathcal{R}$ can be thought of as the radius of curvature of the space.\footnote{The $\mathcal{R}_{ijkl}$ and $\mathcal{R}$ should \textit{not} to be confused with the curvature tensor $^3R_{ijkl}$ and the Ricci scalar $^3R$ of the 3-space orthogonal to $u^i$.} The natural generalization of $\nabla^2\phi\propto\rho$ is then given by $\mathcal{R}^{-2}\propto\rho=T_{ab}u^au^b$. Working out the left hand side (see e.g., p. 259 of 
Ref.~\cite{key7}), one finds that
 $
G_{ab}u^au^b=\kappa T_{ab}u^au^b                                  
$
which is exactly \eq{ee2}!
Thus, \eq{ee2} tells you that the square of the radius of curvature of space is proportional to the reciprocal of the energy density, thereby giving a geometrical meaning to the left hand side.\footnote{The combination $G_{ab} u^au^b$ is also closely related to the ADM Hamiltonian in the conventional approach. But this is a \textit{dynamical} interpretation and not a \textit{geometrical} one. As I will argue, we have gone wrong conceptually in thinking of the metric tensor as a fundamental dynamical variable and erecting a structure around this assumption. Also note that these ideas generalize in a simple manner to all \LL\ models of gravity \cite{ll} and are not limited to Einstein's theory. In this article, however, I will concentrate on Einstein gravity.}

\subsection{A guiding principle for dynamics}
\label{subsec:gpd}

As I said, all the three formulations --- based on \eq{ee1}, \eq{ee2} and  \eq{ee3} --- lead to the same algebraic consequences for classical gravity.  That is, once you specify $T_{ab}$ (and $\Lambda$ in case of \eq{ee4}) and solve the resulting differential equations, you will end up with the same spacetime geometry and same observable consequences. This raises the question: 
Is there a physical principle which will allow us to distinguish between \eq{ee1}, \eq{ee2} and \eq{ee3}, selecting one of them as the correct approach? 

There is indeed one which  will be the cornerstone of the approach I describe in this article.
Recall that the equations of motion for matter, derived from an action principle, remain invariant if you add a constant to the matter Lagrangian, \textit{i.e.}, under the change $L_{m} \to L_{m} + $ constant. This encodes the principle that the dynamics is immune to the shift in the the zero level of energy density. Motivated by this fact, it is reasonable to postulate that the gravitational field equations should also preserve this symmetry, which is already present in the matter sector. Since the energy momentum tensor $T_{ab}$ will occur, in one form or another, as the source for gravity (as can be argued from the principle of equivalence and considerations of the Newtonian limit), this leads to the  postulate:

\begin{itemize}
 \item[$\blacktriangleright$] The extremum principle  that determines the spacetime dynamics (and hence the field equations) must remain invariant under the change $T^a_b \to T^a_b + $ (constant) $\delta^a_b$.
\end{itemize}

This principle immediately rules out \eq{ee1} and \eq{ee2} as possible choices for the field equation and selects \eq{ee3} as the correct choice; indeed, \eq{ee3} remains invariant under the shift $T^a_b \to T^a_b + $ (constant) $\delta^a_b$ because $\ell^2 =0$ for a null vector. This is a direct  consequence of the guiding principle which I will\cite{A19,A11} take as my basic postulate.
This principle will turn out to be as powerful in determining the  gravitational dynamics as the principle of equivalence was in determining the gravitational kinematics. I will begin by exploring its consequences for the variational formulation of the field equation in the next section.

\section{Variational principles for gravity}
\label{sec:vpg}

The guiding principle introduced above constrains the nature of variational principle, from which one can obtain the gravitational field equations. We get two key constraints:

First, 
this principle  rules out the possibility of varying the metric tensor $g_{ab}$ in any covariant, local action principle to obtain the field equations! It is easy to prove \cite{C7} that if: (i) the action is obtained by integrating  a local, covariant Lagrangian, with the covariant measure $\sqrt{-g}\, d^4x$ and (ii) the field equations are obtained by varying the metric in an  unrestricted manner\footnote{The second condition rules out unimodular theories and their cousins, in which  the metric is varied keeping $\sqrt{-g}$ fixed;  we lack a sound physical motivation for this approach.}  in the action, then the field equations \textit{cannot} remain invariant under $T^a_b \to T^a_b + $ (constant) $\delta^a_b$.  In fact, the shift $L_{m} \to L_{m} + $ constant is no longer a symmetry transformation of the action if the metric is treated as the dynamical variable. So, any variational principle we come up with, cannot have $g_{ab}$ as the dynamical variable. 
You cannot work with the Hilbert action added to the matter action  and vary $g_{ab}$ to get \eq{ee1}. In fact, since \eq{ee1} violates our guiding principle, you don't want to get \eq{ee1} at all. Instead, we are looking for a variational principle which will give us \eq{ee3}.

The second constraint, on any such variational principle leading to \eq{ee3}, is the following: 
Since you cannot introduce $T_{ab}$  by varying $g_{ab}$ in a matter action, the $T_{ab}$ must be present in the functional we vary in some form which does not violate our guiding principle.
The most natural structure, built from $T^a_b$, which maintains the invariance we have demanded, viz. under $T^a_b \to T^a_b + $ (constant) $\delta^a_b$, is given by
\begin{equation}
 \mathcal{H}_m \equiv T_{ab} \ell^a \ell^b
\label{Qtot}
\end{equation} 
where $\ell_a$ is a null vector.\footnote{I want to introduce a minimum number of extra variables to implement the required symmetry. In  $d$-dimensional spacetime, a null vector with $(d-1)$ degrees of freedom is the minimum one needs. For comparison, suppose you introduce, say, a combination $T^{ab}V_{ab}$ with a symmetric traceless tensor $V_{ab}$, in order to maintain the invariance under $T^a_b \to T^a_b + $ (constant) $\delta^a_b$. Then you will  introduce $(1/2)d(d+1)-1$ extra degrees of freedom; in $d=4$, this introduces nine degrees of freedom, which is like introducing three null vectors rather than one.}
This is exactly the combination that appears in the right  hand side of \eq{ee3}. 

The fact that you cannot vary the metric to get the equations of motion can come as a bit of a shock, if you had a traditional upbringing. This can  indeed lead  to trouble if you want to obtain \eq{ee1} but as I said before,  our guiding principle selected  out \eq{ee3} as the correct one. In this equation we have the auxiliary variable $\ell^a$ and one can indeed construct variational principles in which we vary $\ell^a$ and obtain \eq{ee3} and thus \eq{ee4}. So everything is completely consistent within the spirit of the formalism we are developing.

Before proceeding further, let me show you a simple variational principle which satisfies the above criterion and leads to \eq{ee3}. Since you cannot vary the metric, let us consider an action principle \cite{aseemtp} in which we vary a null vector field $\ell^a$. We take the action principle to be:
\begin{equation}
A[\ell, \nabla \ell]= \int \frac{d^4x}{L_P^4} \sqrt{-g}\, \left( L_P^4 T_{ab} \ell^a \ell^b + P^{ab}_{cd} \nabla_a \ell^c\, \nabla_b \ell^d\right)
\label{tpa}
\end{equation} 
where $P^{ab}_{cd}$ is a tensor with the algebraic symmetries of the curvature tensor and is divergence-free in all the indices. We take it to be 
\begin{equation}
P^{ab}_{cd} = \frac{L_P^2}{8\pi} \left( \delta^a_c \delta^b_d - \delta^b_c \delta^a_d \right)
\end{equation} 
It is straightforward to show that varying $\ell^a$  after introducing a Lagrange multiplier to ensure $\ell^2=0$ will lead to the equation $R^i_j - \kappa T^i_j = f(x) \delta^i_j$ which --- in turn --- leads to \eq{ee4}; see footnote 3. So, the action in \eq{tpa} ---
which seems to describe a garden variety null vector field with quadratic coupling --- actually leads to the result you want\footnote{Normally, if you vary a quantity $q_A$ in an extremum principle, you get an evolution equation for $q_A$. Here we vary $\ell_i$ in \eq{tpa} but get an equation constraining the background metric $g_{ab}$! This comes about because, after varying $\ell_i$, we demand that the equation must hold for all $\ell_i$.  While this makes our extremum principle conceptually different from the usual ones, it is perfectly well-defined --- and will make physical sense very soon.} as long as the kinetic energy term has a peculiar structure! 

This algebraic fact can be demystified by noticing that the kinetic energy term in \eq{tpa} can be reduced to the form 
\begin{equation}
P^{ab}_{cd} \nabla_a \ell^c\, \nabla_b \ell^d =  \nabla_a w^a +  \frac{L_P^2}{8\pi} R_{ij}\, \ell^c \ell^j
\end{equation} 
where $w^a = P^{ab}_{cd} \ell^c\nabla_b\ell^d$. So, except for a total divergence which does not contribute to the variation, we are actually working with an action that is proportional to $(R_{ab} - \kappa T_{ab}) \ell^a\ell^b$.  The action does not contain any kinetic energy term for $\ell^a$ at all once you remove the total divergence! Nevertheless, \eq{tpa} is a perfectly legitimate action in which you can vary $\ell^a$ and get the equations we want.\footnote{Incidentally, the full action for matter plus gravity is obtained by adding to $A$ in \eq{tpa} the matter action; i.e., 
$
A_{tot}= A + A_{\rm matter}(\psi_A,g_{ab})
$
where $\psi_A$ denotes the matter variables.
This works with the following extra prescription: You vary $\ell_a$ \textit{first} to get the field equations for gravity, use the on-shell values in the first term $A$ in $A_{tot}$ and extremize  the resulting functional with respect to the matter variables $\psi_A$ to determine the matter equations of motion. In a path integral you integrate over $\ell_a$ first. The reason why you need to vary $\ell_a$ first will become clearer later on, when we identify $\ell_a$ with internal variables describing the spacetime microstructure.}

Since this is a somewhat peculiar situation, I will describe what is going on in a slightly more general context. 
Define 
\begin{equation}
 q[x; \ell_a(x)] \equiv \left(T^a_b (x) - \frac{1}{\kappa} R^a_b (x)\right) \ell_a\ell^b \equiv E^a_b \ell_a\ell^b
\label{tpone}
\end{equation} 
which is a function of $x^i$ through $T^a_b$ and $R^a_b$ and a quadratic \textit{functional} of the null vector field $\ell^a(x)$. Consider now a variational principle based on the expression
\begin{equation}
Q [\ell_a(x)] = \int dV F(q[x;\ell_a])
\label{tptwo}
\end{equation} 
where $F(q)$ is a function of $q$ --- which is, at present, arbitrary --- and $Q$ is treated as a functional of $\ell^a$. 
The $F$ is a scalar and the integration in  \eq{tptwo} is over any (sub)domain of the spacetime with a covariant measure $dV$. (Most of the time, in our later discussion, we will be concerned with an integration over a null surface.)
Consider a variational principle of the form $\delta Q =0$ subject to the constraint that  $\ell^2(x) =0$. 
Incorporating this constraint by a Lagrange multiplier $\lambda(x)$ amounts to changing 
$F(q) \to F(q) + \lambda(x) \delta^a_b \ell_a\ell^b$.
Varying $\ell^b$ and demanding that $\delta Q=0$ for arbitrary $\delta \ell^b$ leads to the condition 
\begin{equation}
\left[ F'(q) E^a_b + \lambda(x) \delta^a_b \right] \ell_a =0; \qquad F'(q) \equiv \frac{dF}{dq}
\label{tpfour}
\end{equation} 
We want the field equations to arise from the demand that the extremum condition $\delta Q =0$ should hold for all $\ell^a$.
For this to work: (i) the expression within the square bracket in \eq{tpfour} should vanish and (ii) $q$, which appears in $F'(q)$, should become independent of $\ell^a$ on-shell. The second condition, in turn, requires 
\begin{equation}
 E^a_b = f(x) \delta^a_b\, ,
\label{tpfive}
\end{equation} 
for some $f(x)$, 
so that $q=0$
on-shell. 
Substituting $E^a_b = f(x)\delta^a_b$ into the square bracket in \eq{tpfour} fixes
 the Lagrange multiplier function $\lambda(x)$ to be 
$ \lambda(x)  = - F'(0) f(x)$
but is otherwise of no consequence.\footnote{Except for the constraint that $F'(0)$ should be finite and non-zero. This does not put any severe condition on the nature of the function $F(q)$. In fact, we will see later that the choice important to us is just $F(q) =q$.} 
Taking the divergence of $E^a_b = f(x)\delta^a_b$ and using the Bianchi identity as well as $\nabla_a T^a_b =0$ determines $f(x)$ to be 
$ f(x) = - (1/\kappa) \left( \Lambda +(1/2) R\right)$ 
where $\Lambda $ is a constant.
Plugging it back into \eq{tpfive}, we get the field equation to be 
\begin{equation}
G^a_b = \kappa\, T^a_b + \Lambda \, \delta^a_b
\end{equation} 
which, of course, is the same as \eq{ee4}. 

Thus, one can introduce a variational principle with an arbitrary function $F(q)$ --- where $q$ is defined by \eq{tpone} --- which will lead to our field equation in \eq{ee3} or \eq{ee4}. We do not have to vary the metric in this approach. The variational principle and the resulting field equation remain invariant under the transformation $T^a_b \to T^a_b + $ (constant)$\delta^a_b$.

Incidentally, the above approach  also leads to a natural quantum theory based on the path integral 
\begin{equation}
Z\equiv \int \mathcal{D} \ell_a \, \delta(\ell^2)   \exp \int dV  F[L_P^4 q]
\label{tp16}
\end{equation} 
where we have used the dimensionless variable $L_P^4 q$.
  The path integral $Z$ in \eq{tp16} is  restricted to null vectors which satisfy the condition $\ell^2=0$. (This is why the path integral is nontrivial even for $F\propto q$ which makes it a Gaussian in $\ell_a$.) 
The classical field equations arise from this expression when: (1) we evaluate it in the saddle point approximation and (2) demand that the result should hold for all $\ell^a$. The first condition is completely standard while the second condition is special to our approach.
But note that $Z = Z[g_{ab}, T_{ab}]$ is a complicated (nonlocal) functional of $g_{ab}$ and $T_{ab}$. Varying $g_{ab}$ in $\ln Z$ will now lead to a nonlocal field equation relating $g_{ab}$ to $T_{ab}$.
  But since the path integral defining $Z$ remains invariant under $T^a_b \to T^a_b + $ (constant)$\delta^a_b$ the extremization of $\ln Z$  will  lead to equations of motion which respects this symmetry.\footnote{To avoid misunderstanding, let me stress that this does not contradict our earlier result, viz. you cannot vary the metric and get equations of motion which are invariant under $T^a_b \to T^a_b + $ (constant)$\delta^a_b$.
That claim is valid only for actions satisfying the locality condition  (i) mentioned in the second para in Sec. \ref{sec:vpg}. The $Z$ here will be a highly  non-local  functional of the metric tensor and $T_{ab}$. The equations resulting from an extremum principle based on $\ln Z$ will obey our guiding principle; but this is not a local variational principle obtained by integrating a scalar Lagrangian over the measure $\sqrt{-g}d^4x$. 
 It is possible that this expression contains information about quantum corrections to the classical gravitational field equations. I hope to describe this model in detail elsewhere.}

\section{Heat density of matter}
\label{sec:hdm}

At this stage the physical meaning of the functional $q[\ell_a(x)]$ --- which depends on the matter sector though the combination  $\mathcal{H}_m$ --- is rather unclear. To understand this, we will first determine the physical meaning of $\mathcal{H}_m$. Since our guiding principle demands that matter enters the variational principle only through this combination $\mathcal{H}_m \equiv T_{ab} \ell^a \ell^b$, it is important to clarify its physical meaning, which --- in turn --- will throw light on the physical meaning of $q$. I will now turn to this task.

To gain some insight, consider first the case of an ideal fluid, with $T^a_b = (\rho+p) u^au_b + p\delta^a_b$.  In this case, the combination $T^a_b \ell_a\ell^b$ is actually the \textit{heat density} $\rho +p = Ts$ where $T$ is the temperature and $s$ is the entropy density of the fluid. (The last equality follows from the Gibbs-Duhem relation. We have chosen the null vector such that $(\ell.u)^2=1$ for simplicity.)  The invariance of $T^a_b \ell_a \ell^b$  under  $T^a_b \to T^a_b + $ (constant)$\delta^a_b$  arises from the fact that the \cc , with the equation of state $p+\rho =0$, has  zero heat density, even though it has non-zero energy density. Our guiding principle --- and \eq{ee3} which is selected out by it --- shows that it is the \textit{heat density} rather than the \textit{energy density} which is the source of gravity. This is the first glimpse of the thermodynamic connection.

But we know that $T^a_bu^bu_a$ is the energy density for \textit{any} kind of $T^a_b$, not just for that of an ideal fluid. How can we interpret $T^a_b \ell_a\ell^b$ as the heat density in a  \textit{general} context when $T^a_b$ could describe any kind of source --- not necessarily a fluid --- for which concepts like temperature or  entropy do not exist intrinsically?  \textit{Remarkably enough, it turns out that you can do this!}. In any spacetime, around any event, we can introduce a class of observers (called  local Rindler observers) who will indeed interpret $T^a_b \ell_a\ell^b$ as the heat density contributed by the matter to a null surface which they perceive as a horizon. This \textit{leads} us to  the concept of local Rindler frames (LRFs) and local Rindler observers, thereby providing us with a thermodynamic interpretation of $T^a_b \ell_a\ell^b$ for any $T^a_b$. Let me describe this in some detail:

\begin{figure}
 \begin{center}
  \includegraphics[scale=0.4]{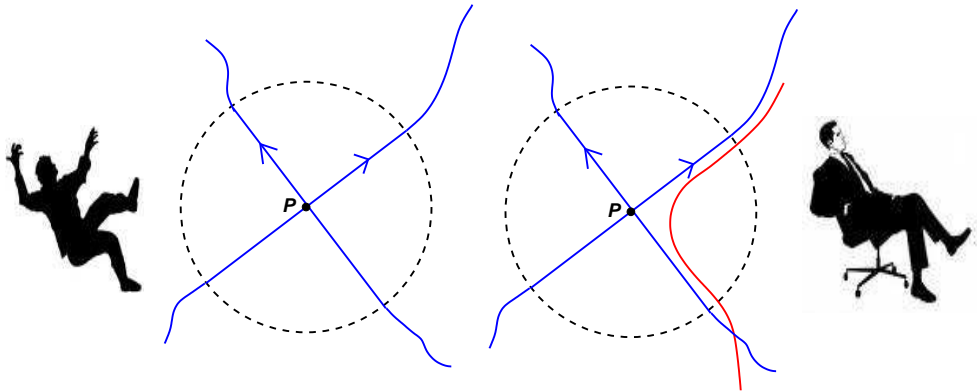}
 \end{center}
\caption{ (a) Left: A freely falling observer and the associated local inertial frame which is well-defined within the region marked by the black circle. The radius of this circle is determined  by the curvature of the spacetime at $\mathcal{P}$.  
Light rays, which travel at 45 degrees in the local inertial frame, define the light cones at $\mathcal{P}$.
(b) Right: A local Rindler observer who is moving with uniform acceleration $a$ in the local inertial frame. For a sufficiently large $a$, his trajectory  will be close to the light cones emanating from $\mathcal{P}$. The light cone will act as a local Rindler horizon to the local Rindler observer who will attribute to it a temperature given by \eq{rindlertemp}. This result arises from the fact that the vacuum fluctuations in the local inertial frame  appear as thermal fluctuations in the local Rindler frame.}
\label{fig:daviesunruh}
\end{figure}
 
We begin by introducing a freely falling frame (FFF) with coordinates $(T, \mathbf{X})$ in a region around some fiducial event $\mathcal{P}$.  Next, we transform from the FFF to a local Rindler frame (LRF; with coordinates $(t,\mathbf{x})$)  through the transformations: $X=\sqrt{2ax}\cosh (at), T=\sqrt{2ax}\sinh (at)$ constructed using some acceleration $a$. (This transformation is for  $X>|T|$ and similar ones exist for other wedges.) One of the null surfaces passing though $\mathcal{P}$, will get mapped to the $X=T$ surface of the FFF and will act as a patch of horizon to the local Rindler observers with the trajectories $x=$ constant [see Fig.~\ref{fig:daviesunruh}].
This construction leads to a very nice result \cite{du1,du2} in quantum field theory. The local vacuum state, defined by the freely-falling observers around $\mathcal{P}$, will appear as a thermal state to the local Rindler observers with a temperature proportional to their acceleration $a$: 
\begin{equation}
 k_BT = \left(\frac{\hbar}{c}\right) \left(\frac{a}{2\pi}\right)
\label{rindlertemp}
\end{equation} 
 (This acceleration $a$ can be related to other geometrical variables of the spacetime in different contexts). 
The existence of the  Davies--Unruh temperature tells us that around \textit{any} event, in \textit{any} spacetime, you will always find a class of observers who will perceive the spacetime as hot.
 
Consider now the flow of energy associated with the matter that crosses the null surface. Nothing strange happens when this is viewed in the FFF by the locally inertial observer. But the local Rindler observer, who attributes a temperature $T$ to the horizon,  views it as a hot surface. Therefore, she will interpret the energy $\Delta E$, dumped on the horizon (by the matter that crosses the null surface in the FFF),  as  energy deposited on a \textit{hot} surface, thereby contributing a \textit{heat} content $\Delta Q=\Delta E$. (Recall that, in the case of, say, a \textit{black hole} horizon, an outside observer will find that any matter takes an infinite amount of time to cross the horizon,  thereby allowing for thermalization to take place. In a similar manner, a local Rindler observer will find that the matter takes a very long time to cross the local Rindler horizon.) It is straightforward to compute $\Delta E$ in terms of $T^a_b$. The LRF provides us with an approximate Killing vector field, $\xi ^{a}$, generating the Lorentz boosts in the FFF, which coincides with a suitably defined\footnote{Since the null vectors have zero norm, there is an overall scaling ambiguity in  expressions involving them. This can be resolved  by considering a family of hyperboloids $\sigma^2\equiv X^2-T^2 =2ax =$ constant and treating the light cone as the degenerate limit $\sigma\to0$ of these hyperboloids. We set $\ell_a= \nabla_a\sigma^2\propto\nabla_a x$ and take the corresponding limit.
 The motivation for this choice will become clearer in the later discussion.} null normal $\ell ^{a}$ at the null surface. The heat current arises from the  energy current $T_{ab}\xi ^{b}$ and the total heat energy dumped on 
the null surface will be:
\begin{align}\label{Paper06_New_11}
 Q_{m}=\int \left(T_{ab}\xi ^{b}\right)d\Sigma ^{a}=\int T_{ab}\xi ^{b}\ell ^{a}\sqrt{\gamma}d^{2}x d\lambda
=\int T_{ab}\ell ^{b}\ell ^{a}\sqrt{\gamma}d^{2}x d\lambda
\end{align}
where we have used the fact that $\xi ^{a} \to \ell ^{a}$ on the null surface. 
Therefore, the combination
\begin{equation}
 \mathcal{H}_m\equiv \frac{dQ_{m}}{\sqrt{\gamma}d^{2}xd\lambda}=T_{ab} \ell^a\ell^b
\label{hmatter} 
\end{equation}
can indeed be interpreted  as the heat density (energy per unit area per unit affine time) of the null surface, contributed by matter crossing a local Rindler horizon. This interpretation is valid in the LRF for any kind of $T^a_b$.  The  need to work with $\mathcal{H}_m$, forced on us by the guiding principle, actually \textit{leads us to} the introduction of local Rindler frames in through which we can interpret $\mathcal{H}_m$ as the heat density.

\section{Heat density of spacetime}
\label{sec:hds}

We saw earlier that a variational principle to obtain the field equations can be built from any functional $F[q]$ of the variable $q$, which --- defined in \eq{tpone} --- can be expressed as:
\begin{equation}
q[x; \ell_a(x)] \equiv \mathcal{H}_m + \mathcal{H}_g;\qquad 
\mathcal{H}_g\equiv - \frac{1}{\kappa} R^a_b\ell_a\ell^b
\label{tpone0}
\end{equation} 
In this approach,  the correct field equations could come from a variational principle based on:
\begin{equation}
Q_{\rm tot}=Q_m+Q_g
=\int \sqrt{\gamma}\, d^2x \, d\lambda\, q[\ell] =\int \sqrt{\gamma}\, d^2x \, d\lambda\,\left(\mathcal{H}_m + \mathcal{H}_g\right)
\label{Qtoty0}
\end{equation}
which corresponds to the simplest choice of $F(q) = q$ in \eq{tptwo}. Further, we saw in the last section that  $\mathcal{H}_m$ can be interpreted as the heating rate per unit area  of the null surface by matter so that $Q_m$ is the matter heat content.
If our ideas are on the right track, then it must be possible  to interpret $\mathcal{H}_g$ as the gravitational contribution to the heating rate (per unit area) of the null surface and $Q_g$ as the gravitational contribution to the heat content. \textit{Remarkably enough, it is is indeed possible to provide such an interpretation}; the term $R_{ab}\ell^a\ell^b$ is related to the ``dissipation without dissipation'' \cite{sanvedtp} of the null surfaces, which arises as follows: 

Construct the standard description of a null surface by introducing the second null vector $k^a$ (with $k^a\ell_a=-1$) and defining the 2-metric on the cross-section of the null surface by $q_{ab}=g_{ab}+k_ak_b+\ell_a\ell_b$. Define the expansion $\theta\equiv\nabla_a\ell^a$ and shear $\sigma_{ab}\equiv \theta_{ab}-(1/2)q_{ab}\theta$ for the null surface where $\theta_{ab}=q^i_aq^j_b\nabla_i\ell_j$. (In this construction, it is simpler to take the null congruence  to be affinely parametrized.) One can then prove that \cite{A19}:
\begin{equation}
-\frac{1}{8\pi L_P^2}R_{ab}\ell^a\ell^b\equiv\mathcal{D}+\frac{1}{8\pi L_P^2}\frac{1}{\sqrt{\gamma}}\frac{d}{d\lambda}(\sqrt{\gamma}\theta)
\label{rai}
\end{equation} 
where
\begin{equation}
 \mathcal{D}\equiv\left[2\eta \sigma_{ab}\sigma^{ab}+\zeta\theta^2\right]
\end{equation} 
is the standard viscous heat generation rate of a fluid with shear and bulk viscous coefficients \cite{A26,A27,membrane}
defined\footnote{The fact that the null fluid has negative bulk viscosity coefficient is well-known in literature,\cite{A26,A27,membrane} especially in the case of black hole membrane paradigm. So we will not pause to discuss its features.} as $\eta=1/16\pi L_P^2,\zeta=-1/16\pi L_P^2$. Ignoring the total divergence term in \eq{rai}, we can identify $\mathcal{H}_g=\mathcal{D}$  and write $Q_{tot}$ as
\begin{equation}
Q_{\rm tot}=\int \sqrt{\gamma}\, d^2x \, d\lambda\, \left(T^a_b \ell_a\ell^b +\mathcal{D}\right)
=\int \sqrt{\gamma}\, d^2x \, d\lambda\, \left(T^a_b \ell_a\ell^b +\left[2\eta \sigma_{ab}\sigma^{ab}+\zeta\theta^2\right]\right)
\label{Qtoty}
\end{equation}
Both terms now have an interpretation of the rate of heating (due to matter or gravity).\footnote{Equation (\ref{rai}) is just a restatement of the Raychaudhuri equation. What is relevant in the extremum principle are the \textit{quadratic} terms in shear and expansion, while the term giving the  change in the cross-sectional area of the congruence is a total divergence and is irrelevant. This tells us that ignoring the quadratic terms of the Raychaudhuri equation can miss a key element of physics\cite{dawood1}.} Our extremum principle can indeed be thought of extremising the rate of production of heat on the null surface.

 Since there are null surfaces passing through any event in the spacetime, we can always find observers who see these surfaces being heated up by the matter crossing them! This is something we do \textit{not} want and gravity comes to the rescue. The contribution to the heating from the microscopic degrees of freedom of the spacetime precisely cancels out $\mathcal{H}_m$ on any null surface on-shell. In fact, this allows us to reinterpret the field equation, expressed as \eq{ee3} as a zero-dissipation principle: $\mathcal{H}_g(\ell)+\mathcal{H}_m(\ell)=0$ 
whenever the integrated boundary term (arising from the total divergence in \eq{rai}) vanishes.\footnote{My use of LRF is \textit{strictly limited} to the purpose of interpreting the quantity $\mathcal{H}_m$. In particular, I do \textit{not} introduce the notion of entropy for the Rindler horizon (as proportional to its area) or work with its variation  etc.}

Let me highlight an important feature  related to the  above variational principle for gravity. In physics,
one encounters two kinds of extremum principles. The first kind involves the extremisation of an \textit{action} $A$ and has its roots in the path integral approach to quantum theory based on the amplitude $\exp(iA/\hbar)$. Given a classical action principle $A(\psi_N)$ based on some dynamical variables $\psi_N$ --- which are varied to get the classical equations of motion --- one could hope to construct a quantum version of the theory using the amplitude $\exp(iA/\hbar)$. The conventional approaches to quantum gravity are based on essentially this philosophy. One takes the classical equations to be \eq{ee1} and the metric to be the dynamical variable which is varied in, say, the Hilbert action.  One is then led to models of quantum gravity in which the spacetime metric (or its variants) become quantum dynamical variables. 
But as I have argued, the correct form of the classical field equation is not \eq{ee1} but  \eq{ee3}. You cannot get this equation by varying the metric in an action principle; in fact, the guiding principle tells you that you cannot treat the metric as a dynamical variable at all in a local action. So the variational principle we are looking at is of a different kind.

Such a different kind of extremum principle also arises in physics but in the context of thermodynamics and statistical mechanics. Here the relevant dynamical equations are obtained by extremising the entropy $S$ or the associated number of degrees of freedom $\Omega$ related to $S$ by $S = \ln \Omega$. 
This thermodynamic interpretation of $\mathcal{H}_m$ tells us that the variational principle one should look for in gravity is of the second kind. I will now show how  a more complete picture emerges when we add the gravity sector to the matter sector.

\section{Atoms of space and their distribution function}
\label{sec:f} 

\subsection{Breaking free: An alternative interpretation}
\label{subsec:free}

So far I have treated $\ell^a(x)$ as an external null vector field in the spacetime and hence a function of the coordinate $x^i$. While this approach proves the existence of suitable variational principles --- which obey our criteria and lead to \eq{ee3} --- this is not  a fundamentally new perspective on gravity. There is, however, a reinterpretation of the variational principle based on $F(q)$ and $\mathcal{H}_m$ which will lead us to the
 objective outlined in Sec. 1, viz., to introduction of the phase space for the atoms of spacetime.
Let me describe this procedure.

The variational principle in \eq{tptwo}, based on $F(q)$, does not contain derivatives of $\ell_a$ and hence is locally algebraic. So it works even if we do not perform the integration over $dV$ in \eq{tptwo}. Let us therefore consider $q(x^i,\ell_a)$ and $F(q)=F(x^i,\ell_a)$ as  \textit{functions} of two independent variables $x^i$ and $\ell^j$ and  think of $\ell^j$ as a set of internal parameters describing the atoms of spacetime at $x^i$. (You can, for example, think of $F(x^i,\ell_a)$ as the distribution function for massless particles in a spacetime with $\ell_a$ being the momenta of the particles.) At any given event, we have a set of different null vectors $\ell_a$ (corresponding, say, to the momenta of different fluid particles). We now demand that $F(x^i,\ell_a)$ should be an extremum when we vary $\ell_a$, subject to the constraint $\ell^2=0$. At any event $P$, this will lead to the condition
 \begin{equation}
\frac{\partial}{\partial \ell^a}[F+\lambda(P) \ell^2] = 0                                                 
\end{equation} 
with the Lagrange multiplier depending on the event $P$. This, of course, immediately leads to \eq{tpfour} and the rest of the results follow. 

In this interpretation, which is better suited for our purpose, we treat
 $\ell_a$ and $x^i$ as independent variables, treating $\ell_a$ as describing some internal spacetime degrees of freedom at $x^i$.   We impose the condition $\ell^2=0$ at each point in spacetime on these internal variables. This is exactly similar in spirit, to the description of a bunch of massless particles using a distribution function $f(x^i,p^j)$ and imposing the condition $p^2=0$ on their momenta. 
 We shall see  that treating $x^i$ and $\ell^j$ as independent variables provides deeper insights.

\subsection{Degrees of freedom for matter}
\label{subsec:matter}

Let us next apply this interpretation to the heat density of matter and treat $\mathcal{H}_m(x,\ell)=T^{ab}(x)\ell_a\ell_b$ 
as a function of two phase space variables $(x^i,\ell_a)$. This
 will allow us to re-express our results in a different form, in terms of the effective number of degrees of freedom. 

Recall that the entropy $S_m$ associated with the heat $Q_m$ in \eq{Paper06_New_11} is given by
$S_m = Q_m/T_H$ where $T_H$ is a temperature introduced essentially for dimensional purposes. (One could, for example, take it to be the temperature associated with the acceleration of the Rindler observers. But, as to be expected, none of the results will depend on its numerical value.) We have
\begin{equation}
S_m(\ell) = \frac{1}{T_H} \int d\lambda\, d^2 x \sqrt{\gamma}\, \mathcal{H}_m(x,\ell) 
= \mu \int \frac{d\lambda\, d^2 x \sqrt{\gamma}}{L_P^3} \, \left(L_P^4 \mathcal{H}_m(x,\ell)\right)
\label{tpnine}
\end{equation}  
where we have introduced suitable factors of $L_P$ to exhibit the dimensionless nature of $S_m$ and defined $\mu \equiv (1/L_PT_H)$.
Replacing the integration by a summation over the relevant spacetime events for conceptual clarity, we can write 
\begin{equation}
S_m(\ell) = \sum_x L_P^4 \mathcal{H}_m (x,\ell) \equiv \sum_x \ln \rho_m(x,\ell) = \ln \prod_x  \rho_m(x,\ell)
\label{tpten}
\end{equation} 
 This introduces the effective number of degrees of matter $\rho_m(x,\ell)=\exp(L_P^4 \mathcal{H}_m (x,\ell))$ at the event $x$, with internal variables $\ell_a$, such that the total number of degrees of freedom associated with the entropy $S_m(\ell)$ is correctly given by
\begin{equation}
\Omega_m(\ell) = \exp S_m(\ell) = \prod_x \rho_m = \prod_x \exp(L_P^4 \mathcal{H}_m ) = \exp\left[\mu \int \frac{d\lambda\, d^2 x \sqrt{\gamma}}{L_P^3} \, \left(L_P^4 \mathcal{H}_m\right)\right]
\label{tpeleven}
\end{equation} 
The first equality is the standard relation between entropy and degrees of freedom, the second expresses the result as a product over the degrees of freedom associated with each event and the third equality expresses it in terms of the variable $\mathcal{H}_m = T^{ab}\ell_a\ell_b$.

This connects up with the discussion in Sec. \ref{sec:grfratoms}   and, in particular, with the result quoted  in the first equation in  \eq{rt4}. There are, however, two points to be noted with this identification. First, recall that $S_m$ in \eq{tpnine} was defined for a fiducial null surface with the normal $\ell^a$, which is a function of coordinates (that is, $\ell^a = \ell^a (x^i)$) and depends on the null surface. Even at this stage, we could consider \textit{different} null surfaces passing through a given event $x^i$, each leading to a different null vector $\ell_a$ at that event. We have formalized this independence of $x^i$ and $\ell_a$ in defining $\rho_m(x,\ell)$ by treating $x^i$ and $\ell^a$ as two independent vectors. 
This is  consistent with the manner 
in which $\rho_m$ and $\Omega_m$ were introduced in Sec. \ref{sec:grfratoms}. At this stage, we are only concerned with the algebraic form of $\rho_m$ and its relation to $T_{ab}$. Later on, I will describe how $\ell^a$ actually arises as a variable describing the \md\ at an event. 

The second point has to do with the observer dependence of entropy as well as the degrees of freedom which contribute to the entropy. There is a folklore belief that the degrees of freedom have some kind of absolute reality and are observer independent. This is, of course, not true when we take into account the thermodynamics of horizons. The temperature and the entropy associated with the local Rindler horizon (or, for that matter, the black hole horizon) is totally observer dependent. Since $S=\ln \Omega$, we reach the conclusion that the degrees of freedom involved in the definition of horizon entropy are also observer dependent. This curious feature arises mathematically from the following facts \cite{tpbibhas1}: In general, one can eliminate the gauge degrees of freedom of spacetime through the diffeomorphisms $x^a\to x^a+\xi^a(x)$. But an observer who perceives a null surface as a horizon can only invoke a subset  of all diffeomorphisms viz., those which  preserve the null surface as a horizon. This means that such an observer can eliminate only a subset of all the degrees of freedom using the diffeomorphisms available to her. This makes certain gauge degrees of freedom  appear as physical degrees of freedom\cite{tpbibhas1} for such an observer thereby leading to non-zero entropy. This fact is not directly relevant to our discussion but I mention it only because of the prevalent misconception that the degrees of freedom are observer independent.

\subsection{Degrees of freedom of quantum geometry}
\label{subsec:geometry}

Based on the ideas described above, I will now look for an extremum principle which maximizes the total number of degrees of freedom of geometry plus matter. This will involve extremising an expression of the kind 
\begin{equation}
\Omega_{\rm tot} = \prod_\ell \Omega_g(\ell) \Omega_m(\ell) \equiv \prod_\ell\ \prod_x \rho_g(x,\ell)\,  \rho_m(x,\ell)
\label{Omtot1} 
\end{equation} 
where $\Omega_m(\ell)$ is defined in \eq{tpeleven} above and $\Omega_g$ is the microscopic degrees of freedom of spacetime geometry, which is the number of atoms of space\footnote{In the case of normal fluid in spacetime, the integral of the distribution function $f(x,p)$ over $p_a$ gives the number density of particles at $x^i$. One can also integrate $f(x,p)$ over all space and get the number density of particles with momentum $p$. We usually do not compute this quantity since it is not very useful in standard fluid mechanics.} with an internal variable $\ell_a$.

The expression for $\rho_m$ in \eq{tpten} was originally obtained from an integral over a null surface in \eq{tpnine}. In that context, $\ell^a(x)$ is a vector field defined on the null surface. But  --- as I stressed earlier --- we could also think of $\ell^a$ as an additional vector  independent of $x^i$ in the expression for $\rho_m$. With such an interpretation, $\rho_g(x,\ell)$ becomes completely analogous to a distribution function with $x^i$ and $\ell^i$ denoting independent phase space variables. We shall adopt this interpretation in what follows since it offers a better insight into the \md.

Exponentiating this expression with respect to the product over $x$ and converting the sum using the measure introduced earlier in going from \eq{tpnine} to \eq{tpten}, we find that 
\begin{equation}
\Omega_{\rm tot} =  \prod_\ell\exp \sum_x \left( \ln \rho_g + L_P^4 \mathcal{H}_m\right) =\prod_\ell \exp\mu \int \frac{d\lambda\, d^2 x \sqrt{\gamma}}{L_P^3} \,\left( \ln \rho_g + L_P^4 \mathcal{H}_m\right) 
\label{tp13}
\end{equation} 
The maximization of this expression  should give us the classical field equation at length scales much larger than $L_P$. This, in turn, requires us to determine the number density of the   atoms of space, $\rho_g(x,\ell)$ at any given event. 

I will show in the next section how one could determine $\rho_g(x,\ell)$ from microscopic considerations leading to the expression 
\begin{equation}
\rho_g(x,\ell) \cong 1 - \frac{L_P^2}{8\pi}\, R^a_b \ell_a\ell^b + \mathcal{O}(L_P^4 R^2)
\label{tp14}
\end{equation} 
which is correct to the lowest order  we are interested in and is adequate to obtain the classical field equation. Substituting \eq{tp14}
into \eq{tp13} we find that $\Omega_{\rm tot}$ is the product of terms of  the form $\exp(q)$ where $q$ defined in \eq{tptwo}. The variational principle is now based on a $F(q)$ which is linear in $q$:
\begin{equation}
F(q) \cong  - \frac{L_P^2}{8\pi}\, R^a_b \ell_a\ell^b + L_P^4\, T^a_b \ell_a\ell^b
\end{equation} 
As we have demonstrated earlier, such a variational principle correctly leads to \eq{ee3} or \eq{ee4} which is what we are after.
Thus, we have obtained classical gravity from a thermodynamic variational principle maximizing the number of degrees of freedom $\Omega_{\rm tot}$ of matter plus gravity. 
I will now describe how we can obtain the expression for $\rho_g$  used in \eq{tp14}.

\section{Area associated with a spacetime  event}
\label{sec:zpl}

It is natural to assume that the number of atoms of space, $\rho_g$, (\textit{i.e.}, the \md) at an event $\mathcal{P}$ should be proportional to either the area or volume (which are the two most primitive geometrical constructs) that we can ``associate with'' the event $\mathcal{P}$. 
What we need to do is to give a precise meaning to the phrase, ``area or volume associated with'' the event $\mathcal{P}$.

For this task,  I will first introduce the notion of an equi-geodesic surface,  
which can be done either in 
the Euclidean sector or in the Lorentzian sector;  I will work in the Euclidean sector. 
An equi-geodesic surface $\mathcal{S}$ is the set of all points at the same geodesic distance $\sigma$ from some specific point, which we take to be the origin \cite{D1,D4,D5,D6}.  A natural  system of coordinates to describe such a surface is given by $(\sigma, \theta_1, \theta_2, \theta_3)$ where $\sigma$, the geodesic distance from the origin, acts as the ``radial'' coordinate and $\theta_\alpha$ are the angular coordinates on the 
equi-geodesic surfaces corresponding to $\sigma =$ constant. 
The metric in this coordinate system is given by: 
\begin{equation}
 ds^2_E = d\sigma^2 + h_{\alpha\beta} dx^\alpha dx^\beta 
\label{sync}
\end{equation} 
where $h_{\alpha\beta} $ is the  metric\footnote{This is the Euclidean analogue of the synchronous frame in the Lorentzian spacetime, with $x^\alpha$ being the angular coordinates.} induced on $\mathcal{S}$.
The two primitive quantities we can now introduce  are the volume element $\sqrt{g}\, d^4x$ in the bulk, and the area element for $\mathcal{S}$ given by $\sqrt{h}\, d^3x$. For the metric in \eq{sync}, $\sqrt{g} = \sqrt{h}$, and hence, both these  measures are the same. Using standard differential geometry, we can show \cite{D8} that, in the limit of $\sigma \to 0$,  these quantities are given by: 
\begin{align}
\sqrt{h}= \sqrt{g}=\sigma^3\left(1-\frac{1}{6}\mathcal{E}\sigma ^{2}\right)\sqrt{h_\Omega}; 
\quad \mathcal{E}\equiv R^a_bn_an^b
\label{gh}
\end{align}
where $n_a=\nabla_a\sigma$ is the normal to $\mathcal{S}$ and $\sqrt{h_\Omega}$ arises from the standard metric determinant of the angular part of a unit sphere.  The second term involving  $\mathcal{E}$ gives the curvature correction to the area of (or the volume enclosed by) an equi-geodesic surface. This \eq{gh} describes a  standard result in differential geometry and is often presented as a measure of the curvature at any event. 

I can now  ``associate'' an area (or volume) with a point $P$ in a fairly natural way  by the following limiting procedure:
(i) Construct an equi-geodesic surface $\mathcal{S}$ around a point $P$ at a geodesic distance $\sigma$. (ii) Calculate the volume enclosed by $\mathcal{S}$ and the surface area of $\mathcal{S}$. (iii) Take the limit of $\sigma\to0$ to define the area (and volume) associated with the point $P$.

This is a natural and well-defined procedure but,
 as you can readily see from \eq{gh}, these measures  vanish in the limit of $\sigma \to 0$. 
This is, of course,  to be expected. The existence of non-zero \md\ requires some kind of discrete structure in the spacetime; they will indeed vanish if  the spacetime is treated as a continuum all the way. (This is analogous to the fact that you can't associate a finite number  of molecules of a fluid with an event $P$ if the fluid is treated as a continuum all the way.)  Classical differential geometry, which leads to \eq{gh}, knows nothing about any discrete spacetime structure and hence cannot give you a nonzero $\rho_g$. To obtain a nonzero $\rho_g$ from the above considerations, we need to ask how the geodesic interval and the spacetime metric get modified in a quantum description of spacetime. We would expect that such a modified description will have a $\sqrt{h}$ (or $\sqrt{g}$) which does not vanish in the coincidence limit. I will now turn to the task of describing a spacetime metric which is modified by quantum gravitational effects, without adhering to any specific quantum gravity model.

There is a significant amount of evidence (see e.g., \cite{D2a,D2b,D2c,D2d,D2e,D2f}) which suggests that a primary effect of quantum gravity will be to introduce into the spacetime  a zero-point length, by modifying the geodesic interval $\sigma^2(x,x')$ between any two events $x$ and $x'$  to a form like $\sigma^2 \to \sigma^2 + L_0^2$ where $L_0$ is a length scale of the order of Planck length.\footnote{A more general modification will take the form of $\sigma^2 \to S(\sigma^2)$  where the function $S(\sigma^2)$ satisfies the constraint $S(0) = L_0^2$ with $S'(0)$ finite. The results I describe here   are  insensitive to the explicit functional form of  $S(\sigma^2)$. So, for the sake of  illustration, I will use $S(\sigma^2) = \sigma^2 + L_0^2$.} 

 While we do not know how the classical metric is modified by quantum gravity, we get  an indirect  handle on it if we assume that quantum gravity introduces a zero point length into the spacetime. This works as follows:
Just as the original $\sigma^2$ is obtained from the original metric $g_{ab}$, we would expect the geodesic interval $S(\sigma^2)$ which incorporates the effects of quantum gravity  to arise from a corresponding quantum gravity-corrected metric \cite{D1}, which we will call the qmetric $q_{ab}$. Of course, no such local, non-singular $q_{ab}$ can exist because, for any such $q_{ab}$, the resulting geodesic interval will vanish in the coincidence limit,  by definition of the integral. We expect $q_{ab}(x,x')$ to be a bitensor, which will be singular at all events in the coincidence~limit $x\to x'$. The fact that the pair $(q_{ab},S(\sigma^2))$ should satisfy the same relationships as $(g_{ab},\sigma^2)$ is enough to
determine \cite{D4,D5,D7} the form of $q_{ab}$. We can express  $q_{ab}$   in terms of $g_{ab}$ (and its associated geodesic interval $\sigma^2$ as: 
\begin{align}
q_{ab}=Ah_{ab}+ B n_{a}n_{b};\qquad q^{ab}=\frac{1}{A}h^{ab}+\frac{1}{B}n^{a}n^{b}
\label{qab}
\end{align}
with
\begin{align}
B=\frac{\sigma ^{2}}{\sigma ^{2}+L_{0}^{2}};\qquad A=\left(\frac{\Delta}{\Delta _{S}}\right)^{2/D_{1}}\frac{\sigma ^{2}+L_{0}^{2}}{\sigma ^{2}};\qquad n_a=\nabla_a\sigma
\label{defns}
\end{align}
where $D$ is the  spacetime dimension, $D_k\equiv D-k$ and
$\Delta$ is the Van-Vleck determinant related to the geodesic interval $\sigma^2 $ by:
\begin{align}
\Delta (x,x')=\frac{1}{\sqrt{g(x)g(x')}}\textrm{det}\left\lbrace \frac{1}{2}\nabla _{a}^{x}\nabla _{b}^{x'}\sigma ^{2}(x,x') \right\rbrace
\end{align}
The $\Delta_S$ is the corresponding quantity computed by replacing  $\sigma^{2}$  by $S(\sigma^{2})$
(and $g_{ab}$ by $q_{ab}$ in the relevant covariant derivatives)
 in the above definition.

For the purpose of determining $\rho_g$, we have to 
compute the area element ($\sqrt{h}\, d^3 x$) of an equi-geodesic surface and the volume element ($\sqrt{q} \ d^4x$) for the region enclosed by it, 
using the  q-metric.  (For the q-metric in \eq{qab}, resulting from the $g_{ab}$ in \eq{sync}, these two measures will not be equal, because $q_{00} \neq 1$.)
If our ideas are correct,  we should get a non-zero value for $\rho_g$ and there must be a valid mathematical reason to prefer one of these measures over the other. 

The computation is 
straightforward  and  (for \mbox{$S(\sigma^2)=\sigma^2+L_0^2$} in $D=4$, though similar results \cite{paperD,D7} hold in the more general case in $D$ dimensions) leads to:
\begin{align}
\sqrt{q}=\sigma \left(\sigma ^{2}+L_{0}^{2}\right)\left[1-\frac{1}{6}\mathcal{E}\left(\sigma ^{2}+L_{0}^{2}\right)\right]\sqrt{h_\Omega}
\label{qfinal}
\end{align}
and\footnote{This result is nontrivial. You might think that the result in \eq{hfinal} (which is  $\sqrt{h}=A^{3/2}\sqrt{g}$)  arises from the standard result in \eq{gh}, by the simple replacement of $\sigma^2\to(\sigma^2+L_{0}^{2})$. But note that this replacement trick does \textit{not} work for the result in \eq{qfinal} (which is $\sqrt{q}=\sqrt{B}A^{3/2}\sqrt{g}$) due to the $\sqrt{B}=\sigma(\sigma ^{2}+L_{0}^{2})^{-1/2}$ factor which has the limiting form $\sqrt{B}\approx\sigma/L_{0}$ when $\sigma\to0$. This is the reason why  each event has zero volume, but a finite area, associated with it!. Some further insight into this curious feature is provided by the following fact:
The leading order dependence of $\sqrt{q}d\sigma\approx\sigma d\sigma$ leads to the volumes scaling as $\sigma^2$ while the area measure is finite. This, in turn, leads to the result \cite{paperD} that \textit{the effective dimension of the quantum-corrected spacetime becomes $D=2$ close to Planck scales,} independent of the original $D$. Similar results  have been noticed by several people (\cite{z1,z2,z3,z4}; also see \cite{nicolini})
in different, but specific, models of quantum gravity. Our approach leads to this result in a fairly \textit{model-independent} manner.}
\begin{align}
\sqrt{h}=\left(\sigma ^{2}+L_{0}^{2}\right)^{3/2}\left[1-\frac{1}{6}\mathcal{E}\left(\sigma ^{2}+L_{0}^{2}\right)\right]\sqrt{h_\Omega}
\label{hfinal}
\end{align}
When $L_{0}^{2}\to0$, we recover the standard result in \eq{gh}, as expected. Our interest, however, is in the coincidence limit $\sigma^2\to0$ taken at finite $L_0$.
Something remarkable happens when we do this. The volume measure $\sqrt{q}$ vanishes (just as it did for the original metric) but $\sqrt{h}$ has a non-zero limit:
\begin{align}
\sqrt{h}= L_{0}^{3}\left[1-\frac{1}{6}\mathcal{E}L_{0}^{2}\right]\sqrt{h_\Omega}
\label{hlimit}
\end{align}
In other words, the qmetric 
attributes to every point in the spacetime a finite area measure, but a zero volume measure! 
Since $L_0^3\sqrt{h_\Omega}$ is the volume measure of the $\sigma=L_0$ surface, we define \cite{tpentropy} the dimensionless density of the \md, as:
\begin{equation}
\rho_g(x^i,n_a)\equiv \frac{\sqrt{h}}{L_0^3\sqrt{h_\Omega}} =1-\frac{1}{6}\mathcal{E}L_{0}^{2}
=1-\frac{1}{6} L_{0}^{2} R_{ab}n^an^b
\label{denast}
\end{equation}

So far we have been working in the Euclidean sector with $n_a=\nabla_a\sigma$ being the normal to the equi-geodesic surface. The limit $\sigma\to0$ in the Euclidean sector makes the equi-geodesic surface shrink down to the origin. \textit{But, in the Lorentzian sector, this limit leads to the null surface which acts as the local Rindler horizon around the chosen event.} Therefore, in this limit, we can identify $n_a$ with the normal to the null surface $\ell_a$ and express
 $\rho_g(x^i,\ell_a)$ as
\begin{equation}
\rho_g(x^i,\ell_a) 
=1-\frac{1}{6} L_{0}^{2} R_{ab}\ell^a\ell^b
\label{denastx}
\end{equation}
To see this in some detail, let us consider the Euclidean version of the local Rindler frame. There are two ways of extending the null surface and the Rindler observers off the $TX$ plane in the Lorentzian sector. You can extend the null line (the 45 degree line in Fig.~\ref{fig:daviesunruh}) to the null \textit{plane}  $T=X$ in spacetime and similarly extend  the hyperboloid. Alternatively, you can 
extend the null line  to the null \textit{cone} by $R^2-T^2=0$ with $R^2=X^2+Y^2+Z^2$ and the hyperboloid ($R^2-T^2=$ constant) will go `around' the null cone in the Lorentzian spacetime (see the left part of Fig.~\ref{fig:lightcones}). Observers living on this hyperboloid will use their respective (rotated) $X$ axis. If we now analytically continue to the Euclidean sector, the hyperboloid $R^2 - T^2 = \sigma^2$ will become a sphere $R^2 + T_E^2 = \sigma_E^2$ (see the right half of Fig.~\ref{fig:lightcones}). The light cone $R^2 - T^2 =0$, which will go over to $R^2 + T_E^2 =0$,  collapses into the origin. The local Rindler observers, living on the hyperboloid $R^2 - T^2 = \sigma^2$, will perceive local patches of the light cone $R^2 - T^2 =0$ as their horizon (see the left half of Fig.~\ref{fig:lightcones}).
 Clearly, taking the limit $\sigma_E \to 0$ in the Euclidean sector corresponds to approaching the local Rindler horizons in the Lorentzian sector. This is the limit in which the hyperboloid degenerates into the light cones emanating from the event $\mathcal{P}$. The normal $n_a$ to the Euclidean sphere can be identified with the normal to the null surface $\ell_a$. The dependence of $\rho_g$ on $n_a$ in the Euclidean equi-geodesic surface  translates into its dependence on the null normal $\ell_a$ in the Lorentzian sector. 

I will now comment on several features which are noteworthy about this approach and the result:

\begin{figure}[t]
 \begin{center}
  \includegraphics[scale=0.48]{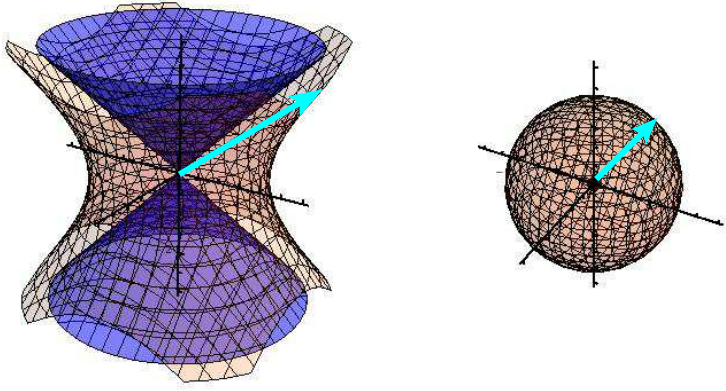}
 \end{center}
\caption{(a) Left: In the local inertial frame, in the Lorentzian sector, the light cones originating from an event (taken to be the origin) are the null surfaces with $R^2-T^2 = 0$  with a normal $\ell_a$. The local Rindler observers who live on the hyperboloid $R^2-T^2 = \sigma^2 = $ constant around these light cones  perceive a patch of the light cone as a local Rindler horizon with a non-zero temperature. The arrow denotes, rather schematically, the normal to the hyperbola. (b) Right: In the Euclidean sector, the hyperboloid $R^2-T^2 = \sigma^2$  maps a sphere $R^2 + T_E^2 = \sigma_E^2$ and the  normal to the hyperboloid  becomes the normal to the sphere. The light cone $R^2 - T^2 =0$ goes over to $R^2 + T_E^2 =0$, and hence collapses into the origin.  The limit $\sigma_E \to 0$, approaching the origin in the Euclidean sector, corresponds to approaching the Rindler horizon in the Lorentzian sector. In this limit,  the hyperboloid degenerates into the two light cones emanating from $\mathcal{P}$. The direction of the normal to the sphere becomes ill-defined in the Euclidean sector, when the radius of the sphere tends to zero. In the Lorentzian sector, we can map it to  the normal to the null surface in the limit when the hyperboloid degenerates to the light cone.
 The dependence of $\rho_g$ on the normal $n_a$ to the Euclidean equi-geodesic surface is what translates into its dependence on the null normal $\ell_a$ in the Lorentzian sector.}
\label{fig:lightcones}
\end{figure}

(i) Our extremum principle in \eq{tp13}, to the leading order, depends on the combination $(\ln \rho_g +L_P^4 T_{ab}\ell_a\ell_b)$. We have defined the the density of \md\ $\rho_g(x,n_a)$ through the limit:
\begin{equation}
\rho_g(x^i,n_a)\equiv \lim_{\sigma\to 0}  \frac{\sqrt{h(x,\sigma)}}{L_0^3\sqrt{h_\Omega}}
\label{denast1}
\end{equation}
in a quantum-corrected spacetime with a zero-point length. This expression had the combination $R_a^bn_bn^a$, at the relevant order, which is crucial. Further, this term comes with a \textit{minus sign}  without which the programme would have failed. 

(ii) On-shell, we have a cancellation between $\ln \rho_g$ and $\ln \rho_m$ so that the total degrees of freedom $\rho_g\rho_m$ becomes unity. This, in turn, implies that the number of degrees of freedom in sphere of radius $R$ is $4\pi R^2/L_P^2$ and a sphere of radius $L_P$ contains $4\pi$ degrees of freedom. We shall have occasion to use this result later on.

(iii) The approach brings to the center-stage the geodesic interval $\sigma^2(x,x')$ (rather than the metric) as the proper variable to describe spacetime geometry \cite{D7}. In a classical spacetime, both $\sigma^2(x,x')$ and $g_{ab}(x)$ contain the same amount of information and each is derivable from the other. But the geodesic interval $\sigma^2(x,x')$ seems to be better suited to take into account quantum gravitational effects to a certain extent.

(iv)  The  spacetime geometry and matter couple to $\ell_a$ through the terms $R^a_b \ell_a \ell^b$ and $T^a_b \ell_a \ell^b$ respectively, thereby leading to an effective coupling between them. The physical nature (and origin) of these  two couplings are quite distinct.  The $T^a_b \ell_a \ell^b$ came from the behaviour of matter crossing the local Rindler horizon and the $\ell_a$ in this expression originally represented the normal to the local Rindler horizon. The $R^a_b \ell_a \ell^b$ term, however, arose from the limit of the area measure $\sqrt{h}$ in a spacetime endowed with a zero-point length. The $n_a$ gets mapped to the normal $\ell_a$ to the null surface through a limiting process when we take the limit $\sigma \to 0$ in the Euclidean sector. This mapping, in turn, depends  on the fact that the condition $\sigma^2 (x,y) =0$ will lead to $x=y$ in the Euclidean space while it will be satisfied by all events connected by a null ray in the Lorentzian space.

(v) Finally, for the sake of completeness, I mention how this formalism can, in principle, be used to obtain semi-classical corrections to the gravitational field equations. One way to do this is to re-write $\Omega_{\rm tot}$ in \eq{Omtot1}, converting both the products into sums\footnote{Note, incidentally, that the double sum in this equation  can be converted to a natural phase space integral of the form 
$
d\Gamma = d^3 V_x d^3 V_n$ with $d^3 V_x= \mu (d\lambda d^2 x \sqrt{\gamma}/L_P^3)$ and  $d^3 V_n= d^4n \delta ( n^2-\epsilon)$ where $\epsilon=1$ in the Euclidean sector and $\epsilon=0$ in the Lorentzian sector.} thereby obtaining 
\begin{equation}
\Omega_{\rm tot}= \exp \sum_x\sum_n \left( \ln \rho_g(\mathcal{G}_N,n_a) + \ln \rho_m(T_{ab}n^an^b)\right)
\end{equation}
Here $\mathcal{G}_N$ denotes different geometrical variables like the metric, curvature tensor etc.
 Performing the summation  over $n_a$, this reduces to the expression 
 \begin{equation}
 \Omega_{\rm tot}= \exp \sum_x S_{\rm eff} (\mathcal{G}_N, T_{ab})
 \label{rt8}
\end{equation}
where we have defined 
\begin{equation}
S_{\rm eff} (\mathcal{G}_N, T_{ab}) = \sum_n \left[ \ln \rho_g ( \mathcal{G}_N, n_a) + \ln \rho_m (T_{ab}n^an^b)\right]
 \label{rt9}
\end{equation}
Extremising $S_{\rm eff}$ with respect to the metric will provide an equation relating the geometrical variables to the energy momentum tensor and will contain corrections to the classical field equation. Moreover, the result will maintain invariance under $T^a_b\to T^a_b+({\rm constant})\delta^a_b$ since the original expression for $\Omega_{\rm tot}$ has this invariance built into it. 

This result, however, is  just formal at this stage because of three reasons: (i) The sum in \eq{rt9} when converted as an integral over $d^3n$ diverges even at the lowest order (where the expression in the square bracket is quadratic in $n^a$). This is to be expected since a formalism analogous to kinetic theory must break down at small scales and one needs to cut off the integration range of the momentum variable $n^a$. It is not clear at this stage how to do this correctly. (ii) We need an  exact expression for $\rho_g(x,n)$ and it is possible to come up with several ansatz for it. (One possibility is to use the result that $\sqrt{h}=1/\Delta$ but there are many geometrical objects which goes over to $\rho_g$ at the leading order.) Even if we come up with a physical criterion, the algebraic expression will be quite complicated and will involve all the spatial derivatives of the curvature tensor. It will be difficult to perform the sum in \eq{rt9} even with a cut-off and then obtain the corrections to classical equations. (iii) Conceptually, I am not happy with varying the metric to get the field equations even from an emergent, effective, action which respects the invariance under $T^a_b\to T^a_b+({\rm constant})\delta^a_b)$,  though it is far better than using the metric as a fundamental dynamical variable.

\section{Cosmological Constant from Cosmic Information}
\label{sec:cc}

The guiding principle I introduced right at the beginning tells you that gravity does not directly couple to the \cc. In the field equations \cc\ arises as an integration constant and --- being a global constant --- needs to be fixed just once. 
We need an extra physical principle for fixing the value of \cc\ and we expect it to arise from the theoretical formalism itself. Indeed it does.

I mentioned earlier that our approach assigns, at the leading order, a single microscopic degree of freedom to each spacetime event on-shell. This means that the quantum gravitational unit of information, associated with a 2-sphere of radius $L_P$ can be taken to be $I_{\rm QG}=4\pi L_P^2/L_P^2=4\pi$. I will now show how this $4\pi$ arises in the study of our universe and helps us to determine the value of the cosmological constant, in a rather intriguing manner \cite{hptpcomment}.

  Let me give you the bottom line first, just to show how intriguing it is. Observations suggest that the evolution of our universe can be described by three different phases, viz., an inflationary phase very early on, followed by a radiation/matter dominated phase which lasted until recently, and an accelerated phase dominated by a small cosmological constant which has started in the near-past and will continue forever. These three phases are characterized by three densities $\rho_{\rm inf}, \rho_{\rm eq}$ (which is the density of matter at the epoch when matter and radiation densities were equal) and $\rho_\Lambda$. 
These three densities make up the signature of our universe, in the sense that the entire evolutionary history can be determined in terms of these numbers.
 Observations determine $\rho_{\rm eq} $ and $\rho_\Lambda$ fairly accurately as: $\rho_{\rm eq} = [(0.86\pm 0.09) \ \text{eV} ]^4$ and 
 $\rho_\Lambda = [(2.26\pm 0.05)\times 10^{-3}  \text{eV}]^4$; we do not have a direct handle on $\rho_{\rm inf}$ but it is usually taken to be about $\rho_{\rm inf}\simeq (10^{15}\ \text{GeV})^4$.
So, in  standard cosmology, these three densities have no relation with each other and they are widely different. 

I now invite you to form a strange dimensionless number $I$ out of these three densities by the definition:
\begin{equation}
I= \frac{1}{9\pi} \, \ln \left( \frac{4}{27} \frac{\rho_{\rm in}^{3/2}}{\rho_\Lambda\,\rho_{\rm eq}^{1/2}}\right) 
\label{strange1}
\end{equation} 
and evaluate its numerical value by plugging in the known values for the three densities. Surprisingly enough, you will find that
\begin{equation}
I   \approx 4\pi \left( 1 \pm \mathcal{O} \left(10^{-3}\right)\right)
\label{strange2}
\end{equation} 
That is, $I = 4\pi$ to an accuracy of one part in thousand for the standard values used in the current cosmological models. This should make you wonder why the right hand side of \eq{strange1} has such a pleasing value as $4\pi$ since it is not often that  such strange things happen. In what follows, I will show  that: (i) the right hand side of \eq{strange1} can actually  be interpreted, in a well-defined manner, as the amount of of cosmic information accessible to an eternal observer and (ii) the reason it is $4\pi$ has to do with the quantum microstructure of spacetime.\footnote{It is an \textit{observational fact}  that $I$ defined via \eq{strange1} has a numerical value $4\pi$ for our universe. You need to be a true believer in coincidences if you think such a result does not tell us anything about our universe and can be completely ignored as ``just one of those things''!} 
Obviously, turning this principle around and taking $I=4\pi$, one can determine the numerical value of \cc\ in terms of the other cosmological parameters, $\rho_{\rm eq}$ and $\rho_{\rm inf}$ which --- eventually --- will be determined from the high energy physics.

\subsection{Accessibility of Cosmic Information}

A key feature of gravity is its ability to control the amount of information accessible to any given observer. Over decades, we have come to realize\cite{info} that information is a physical entity and that anything which affects the flow and accessibility of information will have direct physical consequences. A well-known example of this idea arises in the physics of black holes. It turns out that a similar idea, applied to the cosmos, allows us to solve --- what is usually considered to be --- \textit{the} deepest mystery about our universe, viz., the small numerical value ($\Lambda L_P^2 \approx 10^{-123}$) of the \cc, $\Lambda$. 

Let me begin by recalling how the existence of a non-zero \cc\ prevents an eternal observer $O$ (i.e., an observer whose world line extends to $t\to \infty$ and who makes observations at very late times) from acquiring information from the far reaches of our universe. Let $x(a_2,a_1)$ be the comoving distance traveled by a light signal between the epochs $a=a_1$ and $a=a_2$ with $a_2>a_1$ in the standard FRW model with expansion factor $a(t)$. This is given by:
\begin{equation}
 x(a_2,a_1) = \int_{t_1}^{t_2} \frac{dt}{a(t)} = \int_{a_1}^{a_2} \frac{da}{a^2H(a)}
\end{equation} 
 Therefore the  comoving [$x_\infty(a)$] and proper [$r_\infty (a)$] sizes of the regions of the universe at an epoch $a$, from which $O$ can receive signals at very late times, are given by \cite{fb}:
\begin{equation}
x(\infty,a)\equiv x_\infty(a) = \int_a^\infty \frac{d\bar a}{\bar{a}^2 H(\bar a)}; \qquad r_\infty (a)=ax_\infty(a) \equiv a \int_a^\infty \frac{d\bar a}{\bar{a}^2 H(\bar a)}
\label{one}
\end{equation} 
The nature of $x_\infty(a)$ and $r_\infty (a)$ depends crucially on whether the \cc\ is zero or non-zero.
If $\Lambda =0$ and the universe is dominated by normal matter at late times, then  $H(a) \propto a^{-n}$, with $n>1$ at late times. Then, both these integrals diverge at the \textit{upper} limit as $t\to \infty$, irrespective of the behaviour of the universe at earlier epochs. 
So, in a universe with $\Lambda =0$, the infinite expanse of space will be visible to the eternal observer at late times; there is no blocking of information.

 If $\Lambda \neq 0$ and $H(a) \to H_\Lambda = $ constant at late times, then the situation is quite different. In that case, both the integrals in \eq{one}
are finite at the upper limit and an eternal observer can only access information from a \textit{finite} region of space at an epoch $a$, irrespective of how long she waits.  The amount of accessible Cosmic Information (``CosmIn'') is now reduced from an infinite amount to a finite value, say $I_c$, as a \textit{direct consequence} of the fact that $\Lambda \neq 0$. \textit{It is, therefore, reasonable to expect that the actual numerical value of $\Lambda$ should be related to $I_c$ with $I_c$ decreasing with increasing  $\Lambda$.} I will now derive this relationship. 

 Let us consider a universe (like ours) with three distinct phases of evolution: (i) At very early times, the universe is in a state of inflation with $H(a) = H_{\rm in} = $ constant. (ii) At $a = a_{\rm rh}$, the inflation ends; the universe reheats and becomes radiation-dominated. This goes on till $a=a_{\rm eq}$ which is the epoch of radiation-matter equality. During  $a_{\rm eq} \lesssim a \lesssim a_\Lambda$, the universe is matter-dominated. (iii) For $a \gtrsim a_\Lambda$,  the \cc\ drives the expansion of the universe. I will rescale the expansion factor such that $a_{\rm eq} =1$, and also define the three densities $\rho_\Lambda, \rho_{\rm eq} $ and $\rho_{\rm inf}$ in terms of the corresponding Hubble parameters through the standard relations $\rho_\Lambda = 3 H_\Lambda^2/(8\pi L_P^2)$, etc. 
 The dynamics of such a universe is described by $(\dot a/a)^2 = H_{\rm in}^2 =$ constant during the inflationary phase and by:
\begin{equation}
\left(\frac{\dot a}{a}\right)^2 = H^2(a) = H_\Lambda^2 \left[ 1 + \frac{1}{\sigma^4} \left( \frac{1}{a^4} + \frac{1}{a^3}\right)\right]; \qquad \sigma^4 \equiv \frac{\rho_\Lambda}{\rho_{\rm eq}} \equiv \frac{H^2_\Lambda}{H_{\rm eq}^2}
\end{equation} 
during the post inflationary phase. I will assume instant reheating at $a= a_{\rm rh}$ for simplicity.

The dynamics of our universe is completely determined by three densities $\rho_\Lambda, \rho_{\rm eq} $ and $\rho_{\rm inf}$ which are introduced  as purely empirical parameters. Amongst them, we have some hope that the high energy physics will eventually determine $\rho_{\rm eq} $ and $\rho_{\rm inf}$ in terms of a viable inflationary model and the dark matter content. But we have no theoretical framework which could fix the value of $\rho_\Lambda$ or relate it to the other two densities. The purpose of this section is to relate $\rho_\Lambda$ to $(\rho_{\rm eq},\rho_{\rm inf})$ using the cosmic information content.\footnote{For our universe, observations give $a_{\rm rh}\approx 7.4\times 10^{-25}, a_\Lambda\approx2.8\times 10^3, \sigma\approx 2.6\times 10^{-3},$ if we choose $a_{\rm eq}=1$. We also find that $\rho_{\rm inf}\lesssim (1.94\times 10^{16}\ \text{GeV})^4,
\rho_{\rm eq} = [(0.86\pm 0.09) \ \text{eV} ]^4, 
 \rho_\Lambda = [(2.26\pm 0.05)\times 10^{-3}  \text{eV}]^4$.}

\begin{figure}
\scalebox{0.48}{\input{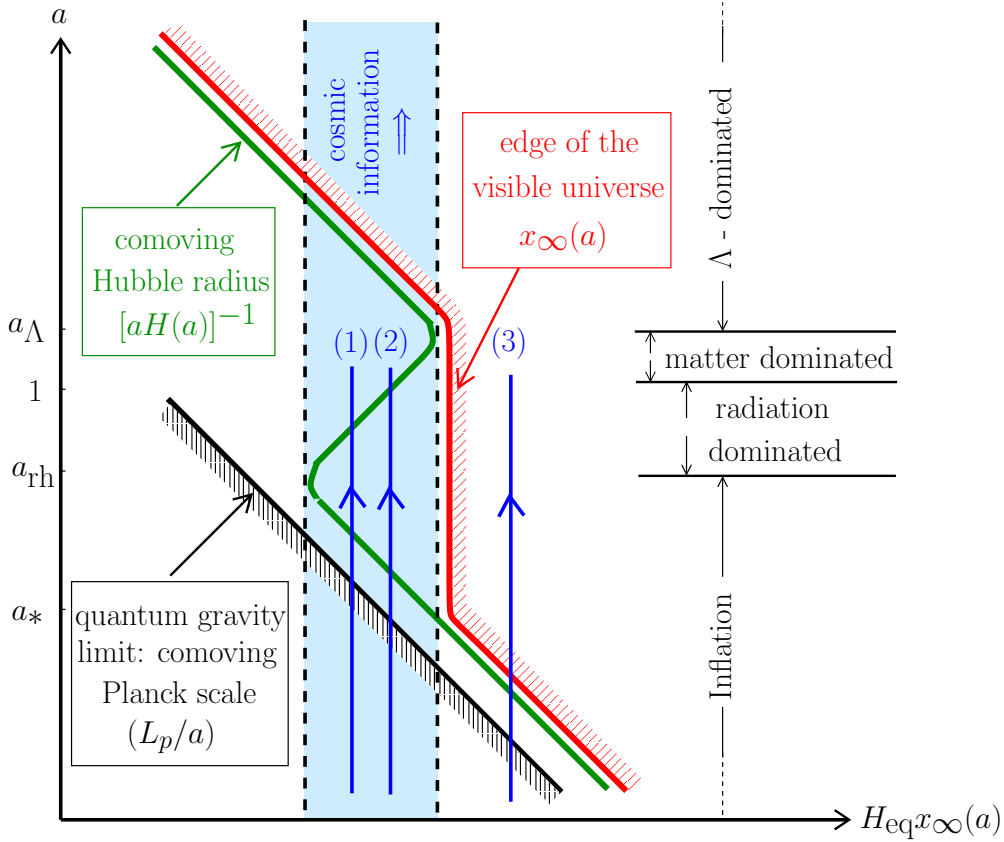}}
 \caption{The various length scales in a universe with an inflationary phase and a non-zero \cc. The red curve gives the maximum comoving size of a region from which signals can reach an observer at very late times. The information in the  shaded region to the right of the red curve is not accessible to an observer even if she waits forever. 
The green curve denotes the comoving Hubble radius. The slanted black curve is the comoving scale corresponding to the Planck length and the shaded region below the black line is dominated by quantum gravitational effects. The vertical lines correspond to different proper length scales which cross the Hubble radius and the horizon. The two lines marked 1 and 2 exit the Hubble radius during inflation and re-enter it during the radiation/matter dominated epoch. These are within the horizon of the observer at the origin (red curve) and are visible to her at, say, $a=a_{\rm rh}$. The line marked 3 corresponds to  a proper length scale which goes out of, not only the Hubble radius, \textit{but also the horizon} and hence will be inaccessible to the observer at, say, $a= a_{\rm rh}$. So the relevant part of  the cosmic information is contained within the blue vertical band, defined by the two vertical lines which are tangential to the comoving Hubble radius at its turning points. The arrows at the top denote the direction of flow of the cosmic information.}
\label{fig:lengthscales}
\end{figure}

The  geometrical features related to $x_\infty(a)$ and other relevant length scales are shown in Fig. \ref{fig:lengthscales}.
  The green curve denotes the \textit{comoving} Hubble radius $d_H(a)/a \equiv 1/ aH(a)$. It  decreases as $1/a$ during the inflationary phase, reaching a minimum at $a= a_{\rm rh}$; it then increases as $a^2$ in the radiation-dominated phase and as $a^{3/2}$ in the matter-dominated phase, reaching a maximum around $a \approx a_\Lambda$;  in the $\Lambda$-dominated phase, it again decreases as $1/a$.
The red curve  gives $x_\infty(a)$  obtained by evaluating the integral in \eq{one} and represents the visibility limit.
During the  $\Lambda$-dominated phase, this curve closely tracks the comoving Hubble radius ($x\approx a_\Lambda^{3/2}/a$) but soon
becomes approximately vertical. During the matter and radiation dominated phases (i.e, during $a_\Lambda \gtrsim a \gtrsim a_{\rm rh}$) the $x_\infty(a)$ is approximately constant --- varying  only by a factor 3 (from $\sim a_\Lambda^{1/2}$ at $a=a_\Lambda$ to $\sim 3 a_\Lambda^{1/2}$ at $a=a_{\rm rh}$) when $a$ varies by a factor $\sim 10^{28}$.  During the inflationary phase, $x_\infty(a)$ again starts tracking $d_H/a$ asymptotically, 
with an approximate behaviour $x_\infty(a) \approx [3 a_\Lambda^{1/2} - a_{\rm rh}] + a_{\rm rh}^2/a$. 
As I said, the region of space from which an eternal observer can acquire information is finite for all finite $a$ if \cc\ is non-zero. 

\subsection{The relation between the \cc\ and   cosmic information}

 Our next task is to quantify the amount of  cosmic information that is accessible to the eternal observer. To do this, recall that a comoving scale  $x= $ constant corresponds to a proper length scale $r = a (t) x$. The proper length scales (e.g., those corresponding to wavelengths of  modes)
will get stretched exponentially during the inflation, and will exit the Hubble radius. They will remain outside the Hubble radius but some of them will re-enter the Hubble radius during the matter/radiation dominated  epoch. (Two such modes are marked as (1) and (2) in Fig. \ref{fig:lengthscales}.)
In contrast, the  mode marked as (3) will exit the Hubble radius but will \textit{never} re-enter it. Such modes actually cross not only the Hubble radius but also the horizon (red line) and become invisible to the eternal observer at, say, the epoch of reheating $a=a_{\rm rh}$. So the modes which are relevant to cosmology are confined to  those between the two dotted horizontal lines which are tangential to the Hubble radius at its turning points. The modes in this blue band contain all the relevant information about our universe and the total number of such modes give us a proper measure of the information content $I_c$. 

I will now estimate how many such  modes cross the Hubble radius during the inflationary phase between $a=a_*$ and $a=a_{\rm rh}$.  
Since the deSitter space is invariant under time translation, the \textit{rate} at which the modes exit the Hubble radius will be a constant. So the number of modes $I(a_2,a_1)$ which cross the Hubble radius during any interval $a_1<a<a_2$ must be proportional to $H(t_2-t_1)$. The total number of modes which cross the Hubble radius during the inflationary epoch will be  proportional to  $N_e\equiv H \Delta t $, where $\Delta t$ is the relevant duration in the inflationary phase. Here
$N_e$ is just the number of e-foldings in the interval $\Delta t $. Therefore,  the CosmIn  is given by:
\begin{equation}
 I_c  \propto N_e
\label{ic1}
\end{equation}
and all we need is the proportionality constant. This can be determined as  
follows:
The number of modes $dN$ in the comoving Hubble volume $V_H(a) = (4\pi/3) (aH)^{-3}$ with wave numbers in the range  $d^3k$ is given by $dN = V_H(a) d^3k/(2\pi)^3=V_H(a)dV_k/(2\pi)^3$ where $dV_k=4\pi k^2 dk$. 
A mode with the comoving wave number $k$ will exit the Hubble radius when $k=k(a)\equiv a H (a)$. So the modes with wave numbers in the range  ($k,k+dk$), where $dk= [d(aH)/da]\, da$, will exit the Hubble radius 
in an interval ($a, a+da$). Hence, the number of modes that cross the Hubble radius during the interval $a_1<a<a_2$ is given by 
\begin{equation}
N(a_2,a_1) = \int_{a_1}^{a_2} \frac{V_H(a)}{(2\pi)^3} \, \frac{dV_k[k(a)]}{da} \, da = \frac{2}{3\pi}\ln \left( \frac{a_2 H_2}{a_1H_1}\right)
\end{equation} 
(Incidentally, this result is applicable for any  $a(t)$.) During inflation, when $a(t) \propto \exp(H_{\rm in} t)$, this expression reduces to $(2/3\pi) \ln(a_2/a_1)$ showing that the proportionality constant in \eq{ic1} is $(2/3\pi)$. Thus the value of CosmIn is given by
\begin{equation}
 I_c = \frac{2}{3\pi} N_e = \frac{2}{3\pi} \ln \left(\frac{a_{\rm rh}}{a_*}\right)
\label{ic2}
\end{equation} 
From the geometry, we can relate the ratio $a_{\rm rh}/a_*$
to the three densities $\rho_\Lambda, \rho_{\rm eq}$ and $\rho_{\rm in}$ which will give $a_{\rm rh}/a_* \propto (\rho_{\rm in}/\rho_{\rm eq})^{1/4} \, (\rho_{\rm eq}/\rho_\Lambda)^{1/6}$. To determine the proportionality constant, we need to evaluate the turning point of the $d_H(a)/a$ curve near $a = a_\Lambda$ which, in turn, requires solving a cubic equation. Doing this \cite{hptp}, we find that the proportionality constant has the value $(4/27)^{1/6}=2^{1/3}/3^{1/2}$. Substituting in to \eq{ic2}, we can achieve our first goal, viz. relating the \textit{non-zero} value of the \cc\ to the \textit{finite} amount of cosmic information accessible to an eternal observer ($I_c$):
\begin{equation}
\rho_\Lambda = \frac{4}{27} \ \frac{\rho_{\rm in}^{3/2}}{\rho_{\rm eq}^{1/2}} \ \exp \left( - 9 \pi I_c\right)
\label{four}
\end{equation} 
As to be expected, the \cc\ vanishes when the information content is infinite ($I_c \to \infty$) vice-versa. 

\subsection{The numerical value of the \cc}

Equation~(\ref{four}) will determine $\rho_\Lambda$ in terms of $\rho_{\rm in}$ and $\rho_{\rm eq}$  if we know the value of CosmIn from some physical consideration. (The  $\rho_{\rm in}$ and $\rho_{\rm eq}$ will be eventually determined from high energy physics in terms of the  inflationary model and the dark matter content of the universe.) 
To determine $I_c$, notice that the  modes which exit the Hubble radius  during the inflationary epoch correspond to sub-Planckian scales in the early part of inflation. In Fig. \ref{fig:lengthscales}, the black line indicates the \textit{comoving} length scale $L_P/a$ corresponding to the Planck length. The region below this line refers to \textit{proper} length scales smaller than the Planck length, and will be dominated by quantum gravitational effects.  The modes which contain the cosmic information cross the comoving Planck length during the earlier stages of evolution and hence will carry the imprint of quantum gravitational effects. So $I_c$ has to be determined by  quantum gravitational considerations.

From our previous discussion, we know that the  unit $I_{\rm QG}$ of quantum gravitational information content of spacetime  is given by the degrees of freedom contained in a 2-sphere of radius $L_P$, viz., 
$I_{\rm QG} = 4\pi L_P^2 / L_P^2 = 4\pi$.
 This suggests that the numerical value for the information content of the cosmos can be taken to be:
\begin{equation}
 I_c = I_{\rm QG} = 4\pi
\end{equation} 
Substituting this into \eq{four}, we get a remarkable formula for the \cc\
\begin{equation}
\rho_\Lambda = \frac{4}{27} \ \frac{\rho_{\rm in}^{3/2}}{\rho_{\rm eq}^{1/2}} \ \exp \left( -36\, \pi^2\right)
\label{five}
\end{equation} 
If we take the typical values $\rho_{\rm in} = (1.2 \times 10^{15}$ GeV)$^4, \rho_{\rm eq} = (0.86$ eV)$^4$, we get $\rho_\Lambda = (2.2 \times 10^{-3}$ eV)$^4$ which agrees well with observed value! 
In other words, the idea that the cosmic information content accessible to an eternal observer, $I_c$, is equal to the basic quantum gravitational unit of information $I_{\rm QG} = 4\pi$, determines the numerical value of the \cc\ correctly.
I will conclude with a few comments: 

\medskip

\noindent (1) The relation $I_c=I_{\rm QG} = 4\pi$, also allows us to determine the relevant number of $e$-foldings in the inflationary epoch which carries the cosmic information. This is given by $N_e = (3\pi/2)I_c = 6\pi^2 \approx 59$, which --- gratifyingly --- leads to an adequate amount of inflation. 

\medskip

\noindent (2) Equation~(\ref{four}) can be reversed to determine the cosmic information content $I_c$ in terms of the three densities. As I mentioned earlier, using the  values for $\rho_\Lambda$ and $\rho_{\rm eq}$ known from observations and taking $\rho_{\rm inf} =( 10^{15}\ \text{GeV})^4$ we find that: 
\begin{equation}
I_c = \frac{1}{9\pi} \, \ln \left( \frac{4}{27} \frac{\rho_{\rm in}^{3/2}}{\rho_\Lambda\,\rho_{\rm eq}^{1/2}}\right) \approx 4\pi \left( 1 \pm \mathcal{O} \left(10^{-3}\right)\right)
\label{six}
\end{equation} 
Thus the current observations  \textit{show that the CosmIn indeed has a value $4\pi$ to the precision of one part in a thousand!}. Because of the logarithmic dependence on the cosmic parameters in \eq{six}, this result is also fairly stable. This renders a purely observational support for the claim $I_c= I_{\rm QG} = 4\pi$.

\medskip

\noindent (3) Theoretically, one would like to determine the value of $\rho_{\Lambda}$ which is the holy grail of cosmology. Observationally, we know  the values of $\rho_{\rm eq}$ and $\rho_\Lambda$ very well today but  have no direct handle on $\rho_{\rm in}$. Using \eq{five}, we can  predict the value of $\rho_{\rm in}$ in terms of the cosmologically determined parameters $\rho_{\rm eq}$ and $\rho_\Lambda$. We then find that  $\rho_{\rm in}^{1/4} = 1.2 \times 10^{15}$ GeV,  which is again a remarkable result.\footnote{In the calculation leading to \eq{six}, I assumed instantaneous reheating; ambiguities in the reheating dynamics can change this result by a factor of few, leading to the prediction $\rho_{\rm in}^{1/4} \approx (1-5) \times 10^{15}$ GeV.} I stress that this is probably the \textit{only} model with quantum gravitational inputs \textit{which leads to a falsifiable prediction}.

\section{Appraisal and discussion}
\label{sec:ad}

There is sufficient amount of evidence to indicate that the correct model for the quantum structure of spacetime will have the following ingredients in one form or the other. I will consider them to be the guiding principles for quantum gravity.

\begin{itemize}
 \item[$\blacktriangleright$] (a) The field equations of classical gravity should emerge as the thermodynamic limit of an underlying statistical mechanics for the \md. This implies that the field equations should come from maximizing a suitably defined density of states.  

 \item[$\blacktriangleright$] (b) The thermodynamics of null surfaces, as well as the observer dependent entropy, which arises from the local loss of information (when a null surface acts as a one-way membrane to a class of observers) should play a key role in  determining the classical limit.  

 \item[$\blacktriangleright$] (c) The gravitational field equation must remain invariant under the transformation $T^a_b \to T^a_b +$ (constant)$\delta^a_b$.

 \item[$\blacktriangleright$] (d) This, in turn, implies that the cosmological constant will arise as an integration constant to the field equations. The cosmic information accessible to an eternal observer, which is rendered finite by non-zero \cc\ must be related to its numerical value. 

 \item[$\blacktriangleright$] (e) A primary effect of quantum gravity should be to modify the classical geodesic interval $\sigma^2(x,x')$ to a function $S(\sigma^2)$ such that $S(0) \equiv L^2_0 $ behaves as the zero point length of the spacetime.

\end{itemize}

The conventional approaches to quantum gravity ignores (b) and (c) completely\footnote{For example,  the temperature and entropy ascribed to a black hole by a geodesic observer inside the event horizon will be different from those ascribed to it by stationary observers at infinity. So if you calculate the entropy of black hole from some observer-independent, microscopic, degrees of freedom in quantum gravity, you are doing something wrong.}, have no clue as to how to handle the \cc\ problem (viz. (d)). They do lead to (e) in some vague sense but do not use the discreteness of spacetime to compute  the density of states and develop (a); the conventional emphasis is on action principles rather than on thermodynamic variational principles.

In this article, I have outlined the procedure which implements all these principles. I started out with (c) to guide us towards the correct form of the classical field equation [viz., \eq{ee3}] and implemented the constraint (c) by introducing the combination $\mathcal{H}_m = T_{ab}\ell^a\ell^b$. The principle (b) allowed us to interpret $\mathcal{H}_m$ in terms of the heating rate of null surfaces in the classical limit and suggested a possible route towards implementing (a). Introducing (e) through the qmetric and calculating the area associated with an event, one could obtain explicit expression for the density of states. Finally, the unit of quantum gravitational information, $I_{\rm QG} = 4\pi$ which was motivated by these considerations, provided a rather surprising solution to the \cc\ problem, 
thereby achieving (d). 

While I expect the principles (a) to (e) to survive all the way to the correct realization of quantum gravity, their implementation may change,  acquiring higher levels of technical and conceptual sophistication. 

The major open question in this approach is the role of matter fields. I have introduced matter through $T_{ab}$ which, at  a fundamental level, is unsatisfactory. This discordance between the ugliness of matter and the beauty of geometry exists even in the conventional formulation of gravity (through, say, $G_{ab} = \kappa T_{ab}$ equating apples to oranges);  our aim is to do better; but we have not succeeded in it.  We cannot vary the metric in an action obtained by integrating a local Lagrangian over $\sqrt{-g} d^4x$, because it will violate principle (c).  As a result, we cannot obtain $T_{ab}$ from the matter action through the variation of the metric. This, by itself, is probably not such a bad deal because, in any case, the description in terms of $T_{ab}$ must break down much before quantum gravitational effects come up. But the problem is that we do not have a prescription which leads to $T_{ab}$ in the field equations once we are forbidden from varying the metric in matter action. The ideas in Sec. \ref{sec:vpg} suggest introducing a null vector field which couples directly to matter and geometry and possibly this idea can be reformulated without explicitly introducing $T_{ab}$. It is rather ironical that the troubles arise from the matter sector rather than from the description of geometry.

\section*{Acknowledgements}

I thank Sumanta Chakraborty, Sunu Engineer, Dawood Kothawala, Kinjalk Lochan and  Hamsa Padmanabhan for discussions and comments on an earlier draft. My research is supported by the J.C. Bose fellowship of DST, India.

\end{document}